\documentclass[twoside]{article}

\usepackage{tabulary,graphicx,times,caption,fancyhdr,amsfonts,amssymb,amsbsy,latexsym,amsmath}
\usepackage[utf8]{inputenc}
\pdfoutput=1
\usepackage{url,multirow,morefloats,floatflt,cancel,tfrupee,textcomp,colortbl,xcolor,pifont}
\usepackage[nointegrals]{wasysym}
\usepackage{graphics}
\usepackage{ulem}
\usepackage{multirow}
\usepackage{float}
\usepackage{inputenc}
\usepackage{xspace}
\usepackage{url}
\usepackage{epstopdf}
\usepackage{amsfonts}
\usepackage{amsmath}
\usepackage{amssymb}
\usepackage{extarrows}
\graphicspath{ {./images/} }
\allowdisplaybreaks

\makeatletter

\usepackage{ifxetex}
\ifxetex\else
  \usepackage{dblfloatfix}
\fi

\@ifundefined{subparagraph}{
\def\subparagraph{\@startsection{paragraph}{5}{2\parindent}{0ex plus 0.1ex minus 0.1ex}%
{0ex}{\normalfont\small\itshape}}%
}{}

\def\URL#1#2{\@ifundefined{href}{#2}{\href{#1}{#2}}}

\def\UrlOrds{\do\*\do\-\do\~\do\'\do\"\do\-}%
\g@addto@macro{\UrlBreaks}{\UrlOrds}

\makeatother

\usepackage[paperheight=11in,paperwidth=8.3in,margin=2.5cm,headsep=.7cm,top=2.5cm]{geometry}
\usepackage[T1]{fontenc}

\renewenvironment{abstract}
	{\trivlist\item[]\leftskip0pt\par\vskip4pt\noindent
  	\textbf{\abstractname}\mbox{\null}\\}
	{\par\noindent\endtrivlist}

\def\keywords#1{\par\medskip\par\noindent\textbf{Keywords}: #1\par}

 \date{} \emergencystretch 8pt

\makeatletter
\def\author#1{\gdef\@author{\hskip-\tabcolsep%
	\parbox{\textwidth}{\raggedright\bfseries#1\\[1pc]}}}
\def\address[#1]#2{\g@addto@macro\@author{\\\hskip-\tabcolsep\parbox{\textwidth}{\raggedright%
	\normalsize\normalfont\textsuperscript{#1}#2}}}
\let\addresslink\textsuperscript
\def\correspondence#1{\g@addto@macro\@author{\\\hskip-\tabcolsep\parbox{\textwidth}{\raggedright%
	\vspace*{10pt}\normalsize\normalfont~\\#1~\\[12pt]}}}
\def\email#1{\g@addto@macro\@author{\\\hskip-\tabcolsep\parbox{\textwidth}{\raggedright%
	\normalsize\normalfont Emails: #1}}}

\def\title#1{\gdef\@title{\vspace*{-30pt}%
	\raggedright\textbf{\@journaltitle}~\\%
  \raggedright\bfseries\ifx\@articleType\@empty\vspace*{20pt}\else%
  \vspace*{20pt}\@articleType\vspace*{20pt}\\\fi#1}}
\let\@journaltitle\@empty \def\journaltitle#1{\gdef\@journaltitle{{\normalfont\itshape#1}}}
\let\@articleType\@empty \def\articletype#1{\gdef\@articleType{{\normalfont\itshape#1}}}

\let\@runningHead\@empty \def\RunningHead#1{\gdef\@runningHead{{\normalfont #1}}}

\usepackage{fancyhdr}
\fancypagestyle{headings}{\fancyhf{}
  \fancyhead[R]{\itshape\@runningHead}
  \fancyfoot[C]{\thepage}}
\fancypagestyle{plain}{%
	\fancyhf{}\fancyhead[R]{Prepared for submission to Hindawi}
  \fancyfoot[C]{\thepage}}
\makeatother

\usepackage[numbers]{natbib}

\usepackage{float,xcolor}

\journaltitle{Advances in High Energy Physics}
\articletype{Research Article} % Research Article/Review Article/Clinical Study

\begin{document}

\title{The fermion-boson map for large $d$ and its connection to lattice transformations}

\author{
		Evangelos G. Filothodoros\addresslink{1}}		
% Affiliation
\address[1]{Institute of Theoretical Physics, Aristotle University of Thessaloniki, Thessaloniki, Greece.}

\correspondence{Correspondence should be addressed to 
    	Evangelos G. Filothodoros; efilotho@physics.auth.gr,vagfil79@gmail.com}

% Emails of author

% Running Head

\maketitle 

% Abstract
\begin{abstract}
I point out that the phase transitions of the $d+1$ Gross-Neveu and $CP^{N-1}$ models at finite temperature and imaginary chemical potential can be mapped to transformations of regular hexagonal and regular triangular lattices to square lattice. The duality elements of two continuous models of fermions and bosons and two discrete lattice models make their appearance offering a new view of their phase transitions. I also show that the fermion-boson map in odd dimensions at finite temperature and imaginary chemical potential has a generalization for arbitrary $d$ that gives an expression of the transfer momentum of fundamental particles that behave like Bloch waves. These particles are travelling inside a periodic potential and scattering from specific surfaces (hexagonal and triangular kind) with a specific ordered construction based on golden ratio formula $\phi=\frac{1}{\phi}+1$ and its generalization. I further argue that this transfer momentum gives us a modified Bragg Law equation which it has a large $d$ limit to the well known expression for the transfer momentum when the scattering lattice is square. Interestingly these surfaces make a family of some first Brillouin zones that interact with particle beams and the maximum amount of momentum of the beam is transferred to them for specific angles related to their construction. Their construction is based on the golden ratio $\phi$ and the Riemann $\zeta(n)$ functions. The zeros and extrema of the Bloch-Wigner-Ramakrishnan $D_d(z)$ functions and Clausen $Cl_d(\theta) $ functions play an important role to the analysis since they allow us not only to study the lattice transformations but also to study the fermionic theory deep inside the strong coupling regime as the dimension of the theory increases. 

% Keywords - if any
\end{abstract}
    
% First level heading
\section{Introduction}
Three-dimensional bosonization physics \cite{Fradkin:1994tt} via  statistical transmutation \cite{Wilczek:1981du,Polyakov:1988md} is a very interesting and recurrent subject in field theory and condensed matter physics, and it has been connected (see e.g. \cite{Karch:2016sxi,Murugan:2016zal,Seiberg:2016gmd,Kachru:2016rui,Meng:2020}) to particle-vortex duality e.g. \cite{Peskin:1977kp}. The extension of fermion-boson duality to finite temperature thermal field theory has also been  considered  (see  for example \cite{Giombi:2011kc,Aharony:2012ns} and references therein), in the context  of various models that describe matter coupled to non-abelian Chern-Simons fields but also the more recent works about exploring the symmetry-breaking of conformal theories in the large-charge limit like \cite{Alvarez}.   Those works present a remarkable progress in our understanding of three-dimensional physics, and its possible holographic higher-spin duals.

Since fermion-boson duality is a fundamental property of three-dimensional quantum physics one may think that it is not necessary to invoke non-abelian gauge fields to make it manifest. Having this idea in mind we revisited in  \cite{Filothodoros:2016txa, Filothodoros:2018}  the finite temperature phase structure of two 3$d$ and odd $d$ systems in general; the fermionic $U(N)$ Gross-Neveu model and the bosonic CP$^{N-1}$ model. We have studied those systems in the canonical formalism by introducing an imaginary chemical potential as a $U(1)$ charge. In such a setup the large-$N$ canonical partition functions are intimately related to the partition functions of the same systems coupled to an abelian Chern-Simons gauge field  expanded around a monopole background in a suitable mean field approximation \cite{Barkeshli:2014ida}. Also, the imaginary $U(1)$ charge density is related to the Chern-Simons level. The generalization of $3d$ theories to arbitrary $odd$ dimensions at finite temperature and imaginary chemical potential is hard due to the non renormalizability of the corresponding models, nevertheless a precise map of their gap equations, free energies and partition functions was demonstrated for a certain region of their phase space \cite{Filothodoros:2016txa, Filothodoros:2018}. The non renormalizability of our models can be tamed by introducing a lattice field-like theory based on Bloch-Wigner-Ramakrishnan functions $D_d(z)$ \cite{Zagier1,Zagier2}.  Our calculations have unveiled the relevance of these functions to the physics of the fermion-boson map and we have argued that there is a non-trivial large $d$ limit of the fermion-boson map which at the level of partition functions is expressed by a general formula with a special characteristic which is the basis for this work.

I begin with a generalization of the Bragg Law of diffraction in Section 2 and I continue with a brief introduction to generic notes about statistical transmutations and the phase transitions of a fermionic theory at imaginary chemical potential in section 3. I continue with a brief review of the three-dimensional and general $odd$ dimensions results of \cite{Filothodoros:2016txa, Filothodoros:2018} in Section 4. In particular, I discuss that when I introduce an imaginary chemical potential the phase structures of the Gross-Neveu and CP$^{N-1}$ models are characterised by the presence of {\it thermal windows} inside which the systems do not have definite fermionic or bosonic properties. The edges of the thermal windows are inflection points of the free energies, while in the middle points of the above windows the systems appear to have completely switched statistics i.e. the fermionic becomes bosonic and vice versa. Encouraged by the odd $d$ analysis I found a remarkable result at the $d\rightarrow\infty$ that points towards  particles that travel inside a lattice \cite{Thesis Filothodoros}. Nevertheless, the edges of the higher-dimensional thermal windows are also inflection points of the free energy and they correspond to zeros and extrema of the $D_d(z)$ functions with $odd$ and even $d$ respectively. This pattern continues for all odd $d$. In previous work we were able to give an analytic formula for the approximate positions of those saddle points in the form of a sequence of rational multiples of $\pi$ that lie between $\pi/3$ and $\pi/2$. As $d\rightarrow\infty$ those points accumulate near $\pi/2$ and the relevant $D_d(e^{i\theta})$ function becomes simply $\sin(\theta)$. This observation allowed us to ask whether the fermion-boson map has a well defined large $d$ limit, and indeed we argued that for models with supersymmetric matter content the partition function duality formula (\ref{susyd}) goes to the simpler one (\ref{limsusy}).

In Section 5 I discuss the main idea of this work where I make the connection of the $U(1)$ charge to a lattice transfer momentum and insert the idea of the hexagons and quadrilaterals conjectures in order to find a correspondence between the generalized thermal windows and the lattice points of specific lattices. In section 6 I examine in detail the setup of the higher dimensional thermal windows for fermions and bosons and the correspondence of the lattices constructions of the conjectures that are the basic idea of the paper. In Section 7 I present a new idea of considering the fermion and boson partition functions as Bloch waves and at Section 8 I discuss the relevance of the strong coupling regime with hexagons and quadrilaterals conjectures which is a new method of lattice regularization. I summarise and offer a few ideas for future work in Section 9. An Appendix contain some technical details and useful formulae for Bloch-Wigner function.

% Second level heading
\section{A generalised Bragg Law}
In physics, Bragg's Law of diffraction is a special case of Laue diffraction, which gives the angles of the scattering from a crystal lattice. When a beam of particles with a wavelength comparable to atomic space between the atoms of a crystal is scattered by a lattice plane, the incidental and the reflected waves remain in phase since the difference between their path lengths is equal to an integer multiple of their wavelength. This path difference between the two waves undergoing interpolation is given by $2d\sin\theta$, where $\theta$ is the scattering angle and $d$ is the interplanar distance.

So, this integer multiple of the wavelength is giving the Bragg's Law as
\begin{equation}
	\label{BragLa}
	n\lambda=2d\sin\theta, n=1,2,3..
\end{equation}

My analysis will focus on the transfer momentum of the scattering on a hexagonal and regular triangle surface of a crystal and its transformations to a square lattice \cite {Brown, Hao}. If we accept that the atoms of the crystal remain fixed, then the change in the incident particle's momentum will be equal to $p_i-p_f$ where $p_i,p_f$ the initial and final momentum respectively. A wave has a momentum $p=\hbar k$ and is a vectorial quantity. The wave number $k$ is the absolute of the wave vector $k=p/\hbar$ and is related to the wavelength $k=2\pi /\lambda$. Frequently, momentum transfer is given in wavenumber units in reciprocal length $Q=k_{f}-k_{i}$. The momentum transfer plays an important role in the evaluation of neutron, X-ray and electron diffraction for the investigation of condensed matter. Bragg diffraction refers to the atomic crystal lattice, conserves the wave energy and thus is called elastic scattering, where the wave numbers of the final and incident particles, $k_f$ and $k_i$, respectively, are equal and since only the direction changes, we calculate a reciprocal lattice vector $G=Q=k_f-k_i$ with the relation to the lattice spacing $G=2\pi/d$. When momentum is conserved, the transfer of momentum corresponds to crystal momentum.

The presentation in $Q$-space is generic and does not depend on the type of the beam and wavelength used but only on the sample system, which allows to compare the results we obtained from  different scattering methods. Then the overall transfer momentum from N particles will be (Figure 1)
\begin{equation}
	\label{Transm}
	Q_{total}=4\pi N\sin\theta/\lambda
\end{equation}

It is obvious that the maximum transfer momentum happens when the angle between the direction of the velocity of the particles and the direction that connects two neighbor atoms horizontally is $\pi/2$. 
 We may write $sin\theta$ like:
 \begin{equation}
 	sin\theta=\left( sin\theta+\lim_{n\to\infty}\sum_{k=2}^{\infty}\frac{sink\theta}{k^n}\right),
 \end{equation}
the lattice space as $d$ and the wavelength as $\lambda$. The above sum goes to zero at the limit where $n\rightarrow \infty$.

\begin{figure}
               \centering
                \includegraphics[scale=0.46]{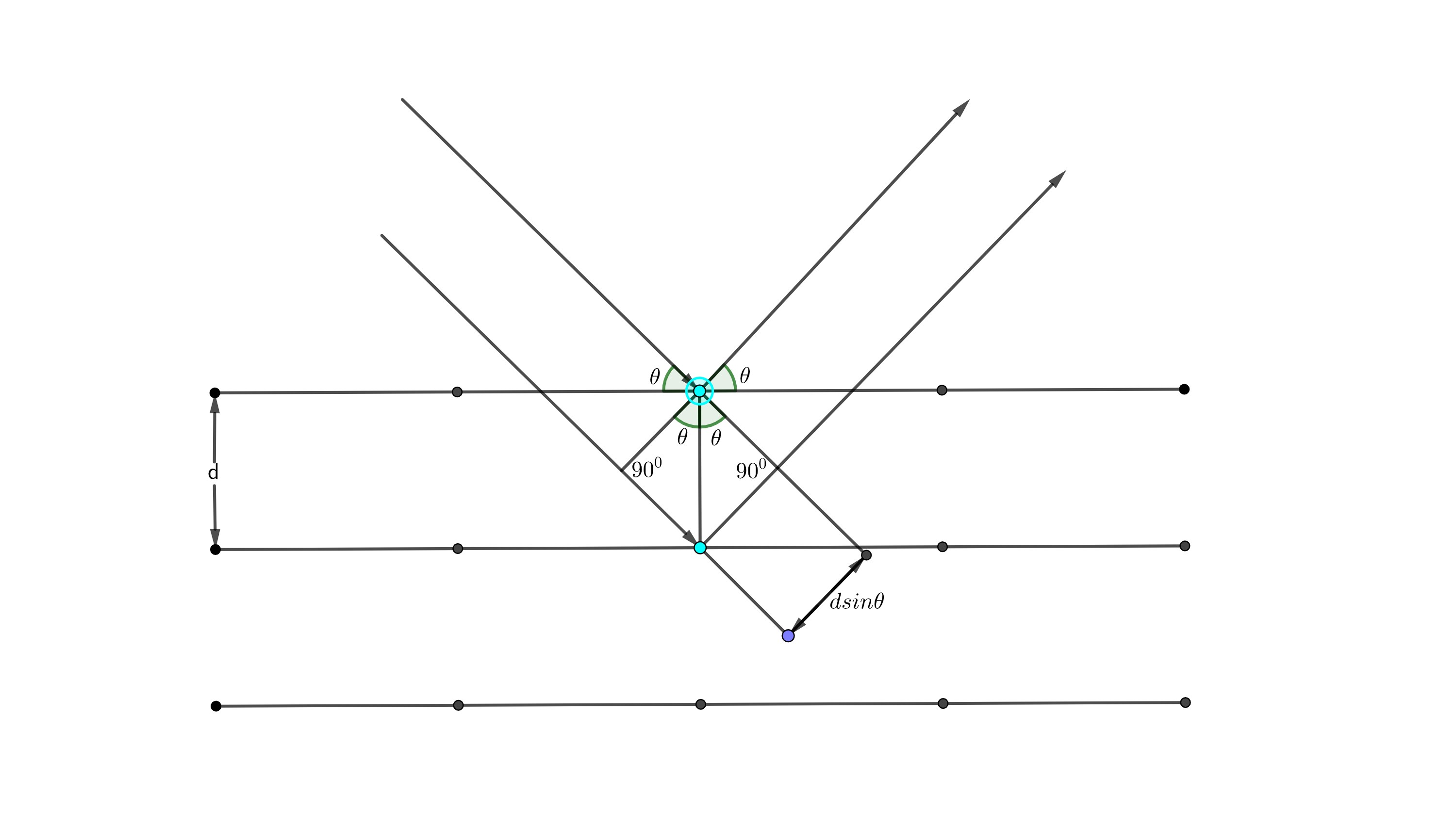}
                 \caption{Bragg diffraction}
\end{figure}

Suppose now that we define a new function $\Omega$ where
\begin{equation}
	\Omega(\theta)\rightarrow \left( sin\theta+\sum_{k=2}^{\infty}\frac{sink\theta}{k^m}\right)
\end{equation}
for arbitrary even $m=2,4,...,\infty$. This $\Omega$ function is Clausen function $Cl_m(\theta)$.	
So the Bragg Law of diffraction of a perturbed lattice with lattice space $d^{'}$ may have a generalised version in arbitrary dimensions:
\begin{equation}
	n\lambda=2d^{'}\Omega(\theta)\rightarrow n\lambda=2d^{'}\left( sin\theta+\sum_{k=2}^{\infty}\frac{sink\theta}{k^m}\right).
\end{equation}
where $dsin\theta=d^{'}\Omega(\theta)$.
The sum in the parenthesis is the Clausen function so:
\begin{equation}
	n\lambda=2d^{'} Cl_m(\theta)	
\end{equation}
The transferred momentum is 
\begin{equation}
	Q^{'}=G
\end{equation}
where $G$ is the reciprocal lattice vector with $G=2\pi/d^{'}$. For $n=1$, $\lambda=1$ and $N$ beam particles we have:
\begin{equation}
	Q^{'}_{total}=4\pi N Cl_m(\theta)
\label{Bragg}
\end{equation}
Since I have argued before that the charge density $Q$ may represent the lattice momentum difference between two wave-functions or the lattice operator that "drives" the incident beam condition to the reflected one, my attention will focus on an equation from our previous work \cite{Filothodoros:2018}.

%\subsection{}

% Third level heading
\section{Statistical transmutation and the $3d$ fermion-boson map at imaginary chemical potential}
I briefly review the relationship of the imaginary chemical potential to statistical transmutation \cite{ZinnJustin:2002ru, SilvaNeto:1998dk} considering for example the fermionic theory. It is easy to see that the presence of the imaginary chemical potential can be cancelled by an abelian gauge transformation for the fermions like:
\begin{align}
\psi(x^0,\bar{x})\mapsto \psi'(x^0,\bar{x})=e^{i\int_0^{x^0} d\tilde{x}^0\alpha_0(\tilde{x}^0)}\psi(x^0,\bar{x})\,,\,\,\\\bar\psi(x^0,\bar{x})\mapsto \bar\psi'(x^0,\bar{x})=e^{-i\int_0^{x^0} d\tilde{x}^0\alpha_0(\tilde{x}^0)}\bar\psi(x^0,\bar{x})\,.
\end{align}
However, at finite temperature the fermions are anti-periodic on the thermal circle
\begin{equation}
\psi(\beta,\bar{x})=-\psi(0,\bar{x})\,,\,\,\,\bar{\psi}(\beta,\bar{x})=-\bar{\psi}(0,\bar{x})\,.
\end{equation}
We then see that the gauge transformed fields would satisfy
\begin{equation}
\psi'(\beta,\bar{x})=-e^{i\theta}\psi'(0,\bar{x})\,,\,\,\,\,\bar\psi'(\beta,\bar{x})=-e^{-i\theta}\bar\psi'(0,\bar{x})\,.
\end{equation}

Hence, the anti-periodic boundary conditions are preserved only if $\theta=2\pi n$, $n\in \mathbb{Z}$. Other values of $\theta$ would "twist" the boundary conditions and change the  statistics of the underlying system. A similar argument goes  through  for  bosonic systems where the complex scalars satisfy periodic boundary conditions on the thermal circle. The twisting of the thermal boundary conditions is the main underlying mechanics behind the possible statistical transmutation in systems whose grand canonical potential is extremized at non trivial values of the imaginary chemical potential.

The calculation of the canonical partition function in systems with  global $U(1)$ charges appears to be agnostic to their underlying microscopic structure e.g. whether the elementary degrees of freedom carrying charge are bosonic or fermionic.  Indeed, it looks like that the only useful piece of information one has is the kind of periodicity of the  partition function, something that could just give a hint regarding the presence of a confinement/deconfinement transition.
 
The situation resembles studies of quantum mechanical systems in a periodic potential like a periodic crystal. If we think of $\theta$ as a periodic coordinate, then is equivalent to the calculation of the overlap between two Bloch wavefunctions that differ by  lattice momentum $Q$ \cite {Blochoverlaps}. Such systems usually generate a band structure which can be studied by restricting the lattice momentum to the first Brillouin zone. Although one generally cannot go very far without using a particular microscopic model at hand, there are certain topological properties of a single band such the eigenvalues of the Zak phase \cite {Zak}, which hold physically relevant information of the system i.e. polarization.  I will pursue  further this line of ideas later and if there is a lesson to be learned is that there may be some universal features of generic quantum systems in periodic potentials which are independent of their fermionic or bosonic microstructure. Hence, a fermion-boson map appears to be generic.  This will be reviewed below by considering two explicit odd dimensional models: the $U(N)$ fermionic Gross-Neveu and the bosonic CP$^{N-1}$  model and the transformations of two lattices, the hexagonal and the triangular.

% Fourth level heading
\section{A brief review of the the $U(N)$ fermionic Gross-Neveu and the bosonic CP$^{N-1}$  models at imaginary chemical potential in odd dimensions}

It is well known that in the absence of a chemical potential $U(N)$ fermionic Gross-Neveu and CP$^{N-1}$ models exhibit very different patterns of symmetry breaking at finite temperature $T$. The Gross-Neveu model has a parity broken phase at low temperatures, which disappears for a critical temperature. On the other hand, while the CP$^{N-1}$ model exhibits the usual continuous symmetry breaking pattern at zero temperature, the broken phase ceases to exist for $T>0$ in accordance to the Mermin-Wagner-Colleman theorem that forbids continuous symmetry breaking at finite $T$ for two-dimensional systems. What instead happens is that when the coupling is tuned to its critical value at $T=0$, then a finite temperature scaling regime with a non-zero thermal mass for the scalars emerges. 
I nevertheless have shown in \cite{Filothodoros:2016txa} that the situation changes in the presence of the imaginary chemical potential, and the corresponding phase structures of the two models can be mapped into each other. I have further observed in \cite{Filothodoros:2016txa}  the relevance of the celebrated Bloch-Wigner function \cite{Zagier1} in our calculations of the gap equations and free energies which for $d$-odd can be expressed as finite sums of Nielsen's generalized polylogarithms \cite{Filothodoros:2016txa, Borwein,Kolbig}, while for $d$-even the corresponding expressions are much more complicated. Moreover, I have already noted the relevance of the 1$d$ theories to the physics of the the 3$d$ models. For these reasons, I will concentrate on studying the generalizations of our 3$d$ models to {\it odd} $d>3$.
This mathematical curiosity, together with the aim to shed more light into the physics of bosonisation, prompted me to study the fermion-boson map in higher dimensions in\cite{Filothodoros:2018} and to extend my thinking to the correlation of the phase transformations of our continuum models with the transformations of specific lattices where they show remarkable similarities and correspondences, with the help of the zeros and maximizations of Bloch-Wigner functions introduced by Zagier. The two models in arbitrary odd dimensions are:

\subsection{The fermions}
The GN model in $d$ Euclidean dimensions is described by the generalization of the 3$d$ action \cite { Filothodoros:2018, Petkou:1998wd, Christiansen:1999uv}
\begin{align}
 S_{GN} = -\int_0^\beta \!\!\!dx^0\int \!\!d^{d-1}\bar{x} \left[\bar{\psi }^{a}(\slash\!\!\!\partial  -i\gamma_0\alpha)\psi ^{a}
+\frac{G_d}{2({\rm Tr}\mathbb I_{d-1})N}\left (\bar{\psi }^{a}\psi ^{a}\right )^{2} +i\alpha NQ_d\right]\,,
\end{align}
with  $Q_d$ the $N$-normalized $d$-dimensional fermionic number density and $a=1,2,..N$.  For odd $d$ we take the dimension of the gamma matrices to be ${\rm Tr}{\mathbb I}_{d-1}=2^{\frac{d-1}{2}}$.

The $d$-dimensional  gap equations become:
\begin{align}
\frac{\sigma_*}{G_d}&=\frac{\sigma_*}{\beta}\sum_{n=-\infty}^\infty\int^\Lambda\!\!\frac{d^{d-1} \bar{p}}{(2\pi)^{d-1}}\frac{1}{\bar{p}^2+(\omega_n-\alpha_*)^2+\sigma_*^2}\,,
\end{align}

\begin{align}
iQ_d&=\lim_{\epsilon\rightarrow 0}\frac{{\rm Tr}\mathbb I_{d-1}}{\beta}\int^\Lambda\!\!\frac{d^{d-1} \bar{p}}{(2\pi)^{d-1}}
\sum_{n=-\infty}^\infty\frac{e^{i\omega_n\epsilon}(\omega_n-\alpha_*)}{\bar{p}^2+(\omega_n-\alpha_*)^2+\sigma_*^2},.
\end{align}

The main issue with the GN model in $d>3$ is that the gap equation has  a finite number of higher order divergent terms as $\Lambda\rightarrow\infty$, which cannot be simply taken care of by the adjustment/renormalization of the single coupling $G_d$ so we have to deal with this issue by using another renormalization method.  On the other hand the charge gap equation is  cut-off independent. Despite these obstructions we were able to extract useful information regarding the phase structure of the model, albeit not as clear cut as in $d=3$. In particular I will exhibit the generalization of the three-dimensional fermion-boson map. 

I will discuss below in some detail the cases $d=5$ and $d=7$ in order to exhibit some of the general features of the higher dimensional models. Starting with $d=5$ and using the results of \cite{Filothodoros:2016txa, Filothodoros:2018}, we have the form of two gap equations as
\begin{align}
\sigma_*\left[-{\cal M}_5\beta^3-D_3(-z_*)-
\frac{1}{2}\ln^2\!|z_*|\left(D_1(-z_*)-\frac{2}{3\pi}\gamma\right)\right]&=0\,,\\
\frac{(2\pi)^2}{{\rm Tr}\mathbb I_{4}}\beta^4Q_5-3i\left[D_4(-z_*)+
\frac{1}{6}\ln^2\!|z_*|D_2(-z_*)\right]&=0\,,
\end{align}
where 
\begin{equation}
\frac{{\cal M}_5}{(2\pi)^2}=\frac{1}{G_{5,*}}-\frac{1}{G_5}\,,\,\,\,\,\gamma=\Lambda\beta\,.
\end{equation}
To derive the first equation we have dropped the infinite number of terms that go as inverse powers of $\Lambda$ and we noted that the last term in parenthesis resembles the corresponding three-dimensional gap equation. Also, in the charge gap equation we could have used the $3d$ case to write it in terms of the charge $Q_3$ of a three-dimensional fermionic model. Remarkably we see signs of a partial deconstruction of the higher dimensional models in terms of lowers dimensional quantities. 

As we go to dimensions $d>3$, the crucial issue is the explicit presence of  the cutoff in the gap equation i.e. compare the $5d$ with the $3d$ case. We emphasised that ${\cal M}_5$ is independent of the cutoff $\Lambda$, hence for a given temperature the $5d$ gap equation is a two-parameter equation for $z_*$. This means that there was no unambiguous way to tune the single coupling constant of the theory, namely the parameter ${\cal M}_5$, in order to obtain a cutoff independent result which is equivalent to the nonrenormalizability of the 5-dimensional theory, unless we use another method. 
The charge gap equation on the other hand had no such issues. 
 
In the presence of an imaginary chemical potential the situation became more interesting, since we encountered again nontrivial zeros of $D_3(-z_*)$ on the unit circle. That means that we were able study the critical theory with ${\cal M}_5=0$. A short excursion in Mathematica  yielded two zeroes for $D_3(-z)$ on the unit circle. Remarkably their positions were approximated to high accuracy by rational multiples of $\pi$ as
\begin{align}
\nonumber D_3(-e^{-i\beta\alpha_*})=Cl_3(\beta\alpha_*\pm\pi)=0\Rightarrow\\ \beta\alpha_*\approx \frac{7\pi}{13}\,{\rm or}\,\beta\alpha_*=\frac{19\pi}{13} \,\,\,({\rm mod}\,2\pi)\,.
\end{align}
Using the periodic properties of the Clausen functions, the relevant results are 
\begin{equation}
Cl_3\left(\frac{6\pi}{13}\right)=Cl_3\left(\frac{20\pi}{13}\right)=0.000362159\,.
\end{equation}
and
\begin{equation}
Q_{5,extr}=\pm i{\rm Tr}\mathbb I_{4}\frac{2}{S_5\beta^4}Cl_4\left(\frac{6\pi}{13}\right)\,, \,\,\,\,\,S_5=\frac{8\pi^2}{3}
\end{equation}
since $Cl_4(\pm 6\pi/13)\approx \pm 0.995777$ are the maximum (minimum) values of $D_4(-z)$ on the unit circle. Notice that $S_5\beta^4$ is the surface of the 4-dimensional sphere. We saw that this patterns generalises to all dimensions.

Finally, when $\beta\alpha_*=\pi$  the gap equation coincides - apart the overall $\sigma_*$ factor - with the corresponding one of the $CP^{N-1}$ that will be given below. The charge is $Q_5=0$ and the system has been bosonized. However, in contrast with the analogous situation in $d=3$, a nonzero solution for $\sigma_*$ in the critical case ${\cal M}_5=0$ depends on the arbitrary parameter $\gamma$. 

I then briefly remember the seven-dimensional case which shows how our results generalized to higher dimensions. The gap equations are
\begin{align}
\sigma_*\left[-{\cal M}_7\beta^5+D_5(-z_*)+\frac{1}{6}\ln^2\!|z_*|\left(D_3(-z_*)+\frac{\gamma^3}{45\pi}\right)
+\frac{1}{24}\ln^4\!|z_*|\left(D_1(-z_*)-\frac{4\gamma}{15\pi}\right)\right]=0\,,\\
\frac{(2\pi)^3}{{\rm Tr}\mathbb I_{6}}\beta^6Q_7+15i\left[D_6(-z_*)+\frac{1}{10}\ln^2\!|z_*|D_4(-z_*)
+\frac{1}{120}\ln^4\!|z_*|D_2(-z_*)\right]=0\,,
\end{align}
where the parameter $\gamma$ has been defined above, and 
\begin{equation}
\frac{3{\cal M}_7}{(2\pi)^3}=\frac{1}{G_{7,*}}-\frac{1}{G_7}\,.
\end{equation}
As before, we clearly see in the first equation of $d=7$ case the presence of terms related to the corresponding three- and five-dimensional  gap equations and, as well as the appearance of the charges $Q_3$ and $Q_5$, through $D_2(-z)$ and $D_4(-z)$,  in the second gap equation.

 Moreover as advertised above we see that it was not possible tuning $\gamma$ to remove the constant terms $D_3(-1)$ and $D_1(-1)$ in the expansion of the gap equation near $\sigma_*=0$, and hence to arrange unambiguously for multicritical behaviour for the effective action. This problem clearly persist for all $d>7$. 

Moving on the non zero chemical potential we can look for zeros of the critical gap equation on the unit circle. Again, their positions are remarkably well approximated, better than in $d=5$, by rational multiples of $\pi$ as
\begin{align}
\nonumber D_5(e^{-i\beta\alpha_*})=Cl_5(\beta\alpha_*\pm \pi)=0\Rightarrow \\
 \beta\alpha_*\approx \frac{26\pi}{51}\,{\rm or}\,\frac{76\pi}{51} \,\,\,({\rm mod}\,2\pi)\,.
\end{align}
The relevant result is 
\begin{equation}
Cl_5\left(\frac{25\pi}{51}\right)=Cl_5\left(\frac{77\pi}{51}\right)=0.000129657\,.
\end{equation}
and we found at these points that
\begin{equation}
Q_{7,extr}=\mp i{\rm Tr}\mathbb I_{6}\frac{2}{S_7\beta^6}Cl_6\left(\frac{25\pi}{51}\right)\,,\,\,\,\,\,S_7=\frac{16\pi^3}{15}\,,
\end{equation}
since $Cl_6(\pm 25\pi/51)\approx \pm 0.999151$ are  the maximum (minimum) values of $D_6(-z)$ on the unit circle. 

The basic features discussed above do not change as we move to higher dimensions. We continue to see the partial deconstruction of the $d$-dimensional gap equations in terms of lower-dimensional pieces. Namely, the $d$ dimensional gap equation contains the $d-2, d-4,...,5,3$-dimensional gap equations,  and the correspondent charge gap equation contains  the $Q_{d-2},Q_{d-4},...,Q_5,Q_3$ charges like

\begin{align}
\beta^{d-1}Q_d+\frac{{\rm Tr}\mathbb I_{d-1}}{2^{d-2}\pi}\beta^{d-3}\ln^2|z_*|Q_{d-2}\nonumber \\
+\frac{{\rm Tr}\mathbb I_{d-1}}{2^{d-1}\pi}\beta^{d-5}\ln^4|z_*|Q_{d-4} \nonumber \\
+\frac{{\rm Tr}\mathbb I_{d-1}}{2^{d}\pi}\beta^{d-7}\ln^6|z_*|Q_{d-6}+\dots\\
= \frac{2 i^d{\rm Tr}\mathbb I_{d-1}}{S_d}D_{d-1}(-z_*) \nonumber
\end{align}
The parameter $\gamma$ appears in the form of an odd polynomial of degree $d-4$, and the condition that $\sigma_*=0$ is an inflection point of the effective action is the $M_{crit}$ equation. 

\subsection{The bosons}
The action of the bosonic theory for general $d$ is a generalization of the $3d$ case \cite{Arefeva:1980ms, DiVecchia, Filothodoros:2018}
\begin{align}
S_{CPN}=\int_0^\beta \!\!\!dx^0 \!\!\int \!\!d^dx\left[|(\partial_0-i\alpha)\phi^a|^2 +|\partial_i\phi|^2  +i\lambda(\bar{\phi}^a\phi^a-\frac{N}{g_d})+iNq_d\alpha\right]\,,\,\,\,a=1,2,..,N\,,
\end{align}
We found the the bosonic formulae for the gap equations and the free energy can be obtained from the corresponding fermionic ones by the identification of the saddle points i.e. $\sigma_*=m_*$ and the shift $z_*\leftrightarrow -z_*$. For example, the gap equations in $d=5$ are
\begin{align}
&-{\cal N}_5\beta^3-D_3\left( z_*\right)-\frac{1}{2}\ln^2|z_*|\left(D_{1}\left( z^*\right)-\frac{2\gamma}{3\pi}\right)=0\,, \\
&\left( 2\pi\right)^2\beta^4q_5+3i\left[D_4\left(z_*\right)+\frac{1}{6}\ln^2|z^*|D_2\left(z_*\right)\right]=0\,,
\end{align}
and the parameter ${\cal N}_5$ by the bosonic version of ${\cal M}_5$. It is clear that the discussion regarding the phase structure of the bosonic models is the shifted image of the corresponding fermionic ones. The higher dimensional results also follow the same pattern.

% Fourth level heading
\section{The charge density $Q$ as a lattice transfer momentum of $N$ particles in odd $d$-The hexagons and quadrilaterals conjectures}
As we have seen before in \cite{Filothodoros:2018} 
\begin{equation}
\label{dual3}
Z_{tot}(\beta\alpha_*)\equiv Z^{(3)}_{f}(\beta\alpha_*+\pi)[Z^{(3)}_b(\beta\alpha_*)]^{\frac{{\rm Tr}{\mathbb I}_{2}}{2}}=e^{i\pi V_2 N \frac{{\rm Tr}{\mathbb I}_{2}}{2}q_3}=e^{N\frac{{\rm Tr}{\mathbb I}_{2}}{2}\frac{V_2}{\beta^2}D_2(z_{*})}\,,
\end{equation}
where $Z^{(3)}_f$ ($Z^{(3)}_b$) denote the $3d$ fermionic (bosonic) canonical partition function. I have kept the ${\rm Tr}{\mathbb I}_{2}$ explicitly in order to compare with the corresponding formula for general $d$ that will be given later. 
The total partition function plays the role of the generating function of all correlation functions. The above equation may be interpreted as giving the difference between the momentum of an incoming and outgoing beam when it scatters from a lattice point and the only change that happens is in the direction. Since the exponential is imaginary (I take the imaginary part of the charge) the above gives us a scattering-like procedure of a Bloch-wave kind beam on a lattice point.
When $\beta\alpha_*=0,\pi$, then $Q_3=q_3=0$ and the above is the well-known statement of fermion/boson duality i.e. the twisted fermionic (namely, imposing period boundary conditions) and the bosonic partition functions are inverse one of the other. However, for $Q_3,q_3\neq 0$ the corresponding partition functions are {\it weighted duals} due to the presence of the real exponential in the r.h.s. of the above equation. We can give an interpretation of that latter weight factor recalling that the gap equations of charge for the Gross-Neveu and the $CP^{N-1}$ models tie $Q_3$  and $q_3$ to the Bloch-Wigner function $D_2(z)$, and hence to the  volume of hyperbolic manifolds. Then the above could be understood as giving the leading "classical" term in a perturbative expansion of a complex Chern-Simons action in inverse powers of the level. Support for such an interpretation also comes from the fact that the extremal values that we have found for the fermionic and bosonic imaginary charges, coincide with the results reported in the study of the partition function of the $SL(2,\mathbb{C})$ CS theory \cite {Witten:1989ip, Gukov:2003na, Gukov:2016njj,Gang:2017hbs} where a complex hyperbolic volume is a combination of a gravity hyperbolic volume together with the Chern-Simons invariant
\begin{equation}
	\label{HypVol}
	\tilde{Vol_M}=Vol_M+iCS_M
\end{equation}
where from my point of view we are dealing with the hypervolume-gravity part when we calculate odd index $D(z)$ functions and a Chern-Simons part when we calculate the even index $D(z)$ functions like the charge $Q$, which is similar with the Chern-Simons part of past works in which theoretically fermionic or bosonic matter was coupled to a Chern-Simons gauge field.
On the other hand, it is also tempting to interpret the above as a "generalized Wilson line" i.e. an overlap between Bloch states at different quasimomenta. The Wilson line operator describes the transport of a Bloch state from quasimomentum $Q$ to $q$. In such a case, the partition functions would correspond to complex conjugate Bloch wave functionals, and the charges $Q_3$ or $q_3$ would play the role of quasimomenta. This hypothesis will be enhanced by the analysis of the next section.  Finally, in previous work we have also suggested a geometric interpretation for this result. Namely, that the bosonic and fermionic free energies  at imaginary chemical potential correspond to partial volumes of an ideal tetrahedron. Their sum gives the entire volume.

Based on the hypothesis I had previously, I can, in line with the generalization of the thermodynamic study of Gross-Neveu and CP$^{N-1}$ theories, give a more general hypothesis on the total particles (that scattered from a lattice) value of the transfer momentum. I have seen that the exponent of the total partition function of a supersymmetric model from charged fermions and bosons gives us the total transfer momentum that occurs when these particles are scattered by a crystal with a hexagonal unit lattice. This is based on the fact that the inner angles of a regular hexagon are exactly $2\pi/3$ so the maximization of the transfer momentum coincides with the maximization of the Clausen function which is equal to the imaginary part of $Li_2(-z)$ on the unit circle. If we look for the corresponding maximization for the charge/momentum at $5,7,9$ .. dimensions we observe that this occurs for specific angles of incidence $7\pi/13$,$26\pi/51$,$103\pi/205$ and so on. These values come from an analytic formula for the approximate positions of the zeros of all $D_{2n-1}(z)$, $n=1,2,..$  functions on the unit circle. We obtain:
\begin{equation}
	D_{2n-1}(e^{-i\beta\alpha_*})\equiv Cl_{2n-1}(\beta\alpha_*)=0\Leftrightarrow \beta\alpha_{*}\approx \theta_n,2\pi -\theta_n\,({\rm mod}\,2\pi)\
\end{equation}
where
\begin{equation}
\label{ThetaN}
	\theta_n=\frac{\pi}{2}\left(1-\frac{5}{4^{n+1}-(-1)^{n+1}}\right)\
\end{equation}
for $n=1,2,3,..$. There is an interesting approximation of these zeros in \cite{Etienne}.

The particles see specific surfaces on Euclidean space with the first Brillouin zones of them including irregular hexagons. At a large $d$ limit the conjecture ends up to the square lattice construction. Somehow as dimension increases the 6th and 5th sides disappear. The conjecture turns  to be as follows:

{\centering
	\begin{tabular}{|c|c|c|c|c|c|c|}
		\hline 
		\multicolumn{7}{|c|}{\textbf{Table 1. The interior angles of hexagons conjecture-Euclidean space}} \\ 
		\hline 
		Dimensions & Angle $1$ & Angle $2$ & Angle $3$ & Angle $4$ & Angle $5$ & Angle $6$ \\ 
		\hline 
		5 & $7\pi/13$ & $7\pi/13$ & $19\pi/26$ & $19\pi/26$ & $19\pi/26$ & 19$\pi$/26\\ 
		\hline 
		7 & $26\pi/51$ & $26\pi/51$ & $38\pi/51$ & $38\pi/51$ & $38\pi/51$ & $38\pi/51$\\ 
		\hline 
		9 & $103\pi/205$ & $103\pi/205$ & $307\pi/410$ & $307\pi/410$ & $307\pi/410$ & $307\pi/410$\\ 
		\hline 
		... & ... & ... & ... & ... & ... & ... \\ 
		\hline 
		$\infty$ & $\pi/2$ & $\pi/2$ & $\pi/2$ & $\pi/2$ & $0$ &$0$ \\ 
		\hline 
	\end{tabular}\par}
Let's see a way to construct these hexagons by using the golden ratio generalized polynomial.

\begin{itemize}
	\item \underline{Figure 2}
\end{itemize} 

$AF=AW=\alpha,WL=WO=\sqrt{2}\alpha$,
$\frac{AW}{sin\omega}=\frac{AL}{sin\theta}=\frac{WL}{sin2\pi/3}\rightarrow \frac{\alpha}{sin\omega}=\frac{AL}{sin\theta}=\frac{WL}{sin2\pi/3}=\frac{2\sqrt6 \alpha}{3}$, so $sin\omega=\frac{\sqrt6}{4}$,
where $\omega$ is the angle between AL and WL and $\theta$ is the angle between AW and WL. 
Also, $sin^2\omega+cos^2\omega=1\rightarrow cos\omega=\frac{\sqrt10}{4}$. But $\theta+\omega=\frac{\pi}{3}\rightarrow$

$sin\theta=sin(\frac{\pi}{3}-\omega)\rightarrow sin\theta=sin\frac{\pi}{3}\cdot cos\omega-cos\frac{\pi}{3}\cdot sin\omega=\frac{\sqrt6}{8}(\sqrt5 -1).$

So, $AL=sin\theta\cdot\frac{2\sqrt6 \alpha}{3}=\frac{(\sqrt5 -1)\alpha}{2}=\frac{\alpha}{\phi}$. At the end we find that
\begin{equation}
	\frac{AF}{AL}=\phi\rightarrow\frac{AL}{AF}=\frac{1}{\phi}
\end{equation}
with the corresponding hexagonal lattice (Figure 2).

\begin{figure}
\centering
\includegraphics[scale=0.34]{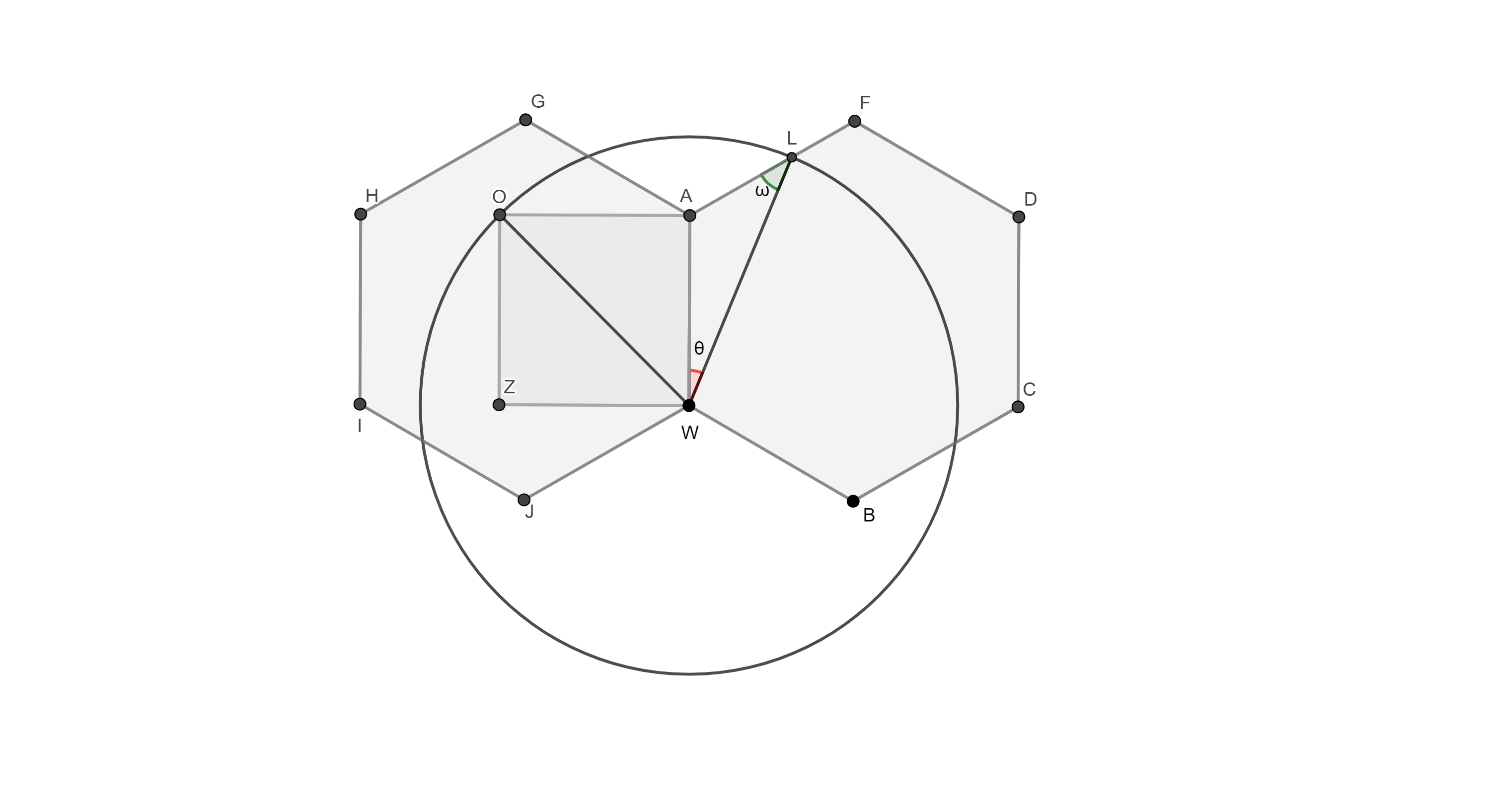}
\caption{Regular hexagonal lattice and golden ratio $\phi$}
\end{figure}

\begin{itemize}
	\item \underline{Figure 3}
\end{itemize}
$AF=AW=\alpha,WL=WO=\sqrt{2}\alpha$

$\frac{AW}{sin\omega}=\frac{AL}{sin\theta}=\frac{WL}{sin7\pi/13}\rightarrow \frac{\alpha}{sin\omega}=\frac{AL}{sin\theta}=\frac{WL}{sin7\pi/13}=1.4246 \alpha$, so $sin\omega=0.70195$.
Also, $sin^2\omega+cos^2\omega=1\rightarrow cos\omega=0.712225$. But $\theta+\omega=\frac{6\pi}{13}\rightarrow$

$sin\theta=sin(\frac{6\pi}{13}-\omega)\rightarrow sin\theta=sin\frac{6\pi}{13}\cdot cos\omega-cos\frac{6\pi}{13}\cdot sin\omega=0.622415.$

So, $AL=sin\theta\cdot 1.4246 \alpha=0.8866925 \alpha$. At the end we find that
\begin{equation}
	\frac{AF}{AL}=1.12778669\approx \zeta(3)\rightarrow\frac{AL}{AF}\approx\frac{1}{\zeta(3)}
\end{equation}
where we have an approximation of $93.41\%$.

\begin{figure}
\centering
\includegraphics[scale=0.38]{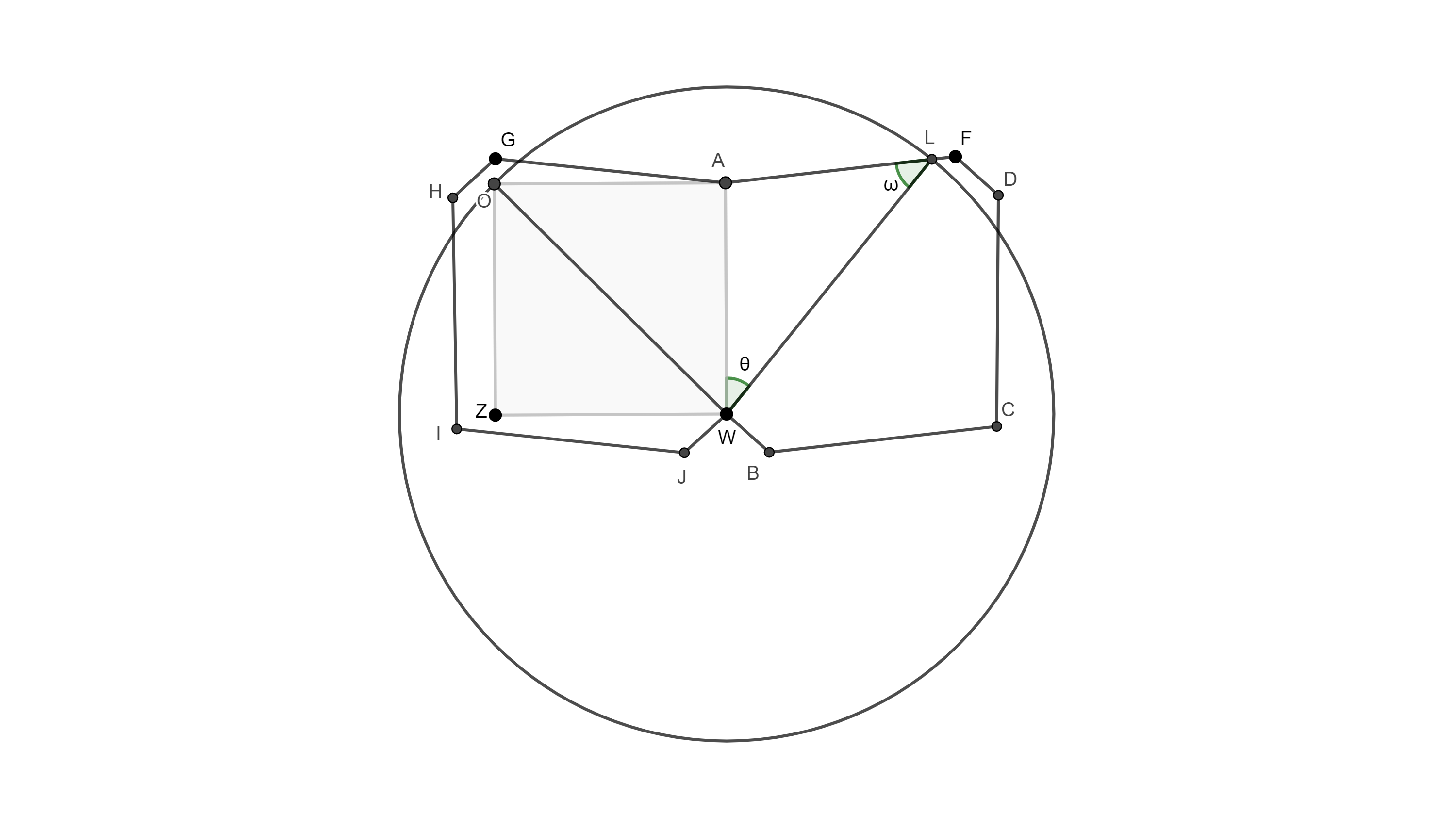}
\caption{Hexagonal lattice and $\zeta(3)$ }
\end{figure}
\begin{itemize}
	\item \underline{Figure 4}
\end{itemize}

$AF=AW=\alpha,WL=WO=\sqrt{2}\alpha$

$\frac{AW}{sin\omega}=\frac{AL}{sin\theta}=\frac{WL}{sin26\pi/51}\rightarrow \frac{\alpha}{sin\omega}=\frac{AL}{sin\theta}=\frac{WL}{sin26\pi/51}=1.414885 \alpha$, so $sin\omega=0.7067712$.
Also, $sin^2\omega+cos^2\omega=1\rightarrow cos\omega=0.7074418$. But $\theta+\omega=\frac{25\pi}{51}\rightarrow$

$sin\theta=sin(\frac{25\pi}{51}-\omega)\rightarrow sin\theta=sin\frac{25\pi}{51}\cdot cos\omega-cos\frac{25\pi}{51}\cdot sin\omega=0.6853407.$

So, $AL=sin\theta\cdot 1.414885\alpha=0.9696782 \alpha$. At the end we find that
\begin{equation}
	\frac{AF}{AL}=1.03126996\approx \zeta(5)\rightarrow\frac{AL}{AF}\approx\frac{1}{\zeta(5)}
\end{equation} 
where now we have a better approximation of $99.45\%$. 
\begin{figure}
\centering
\includegraphics[scale=0.36]{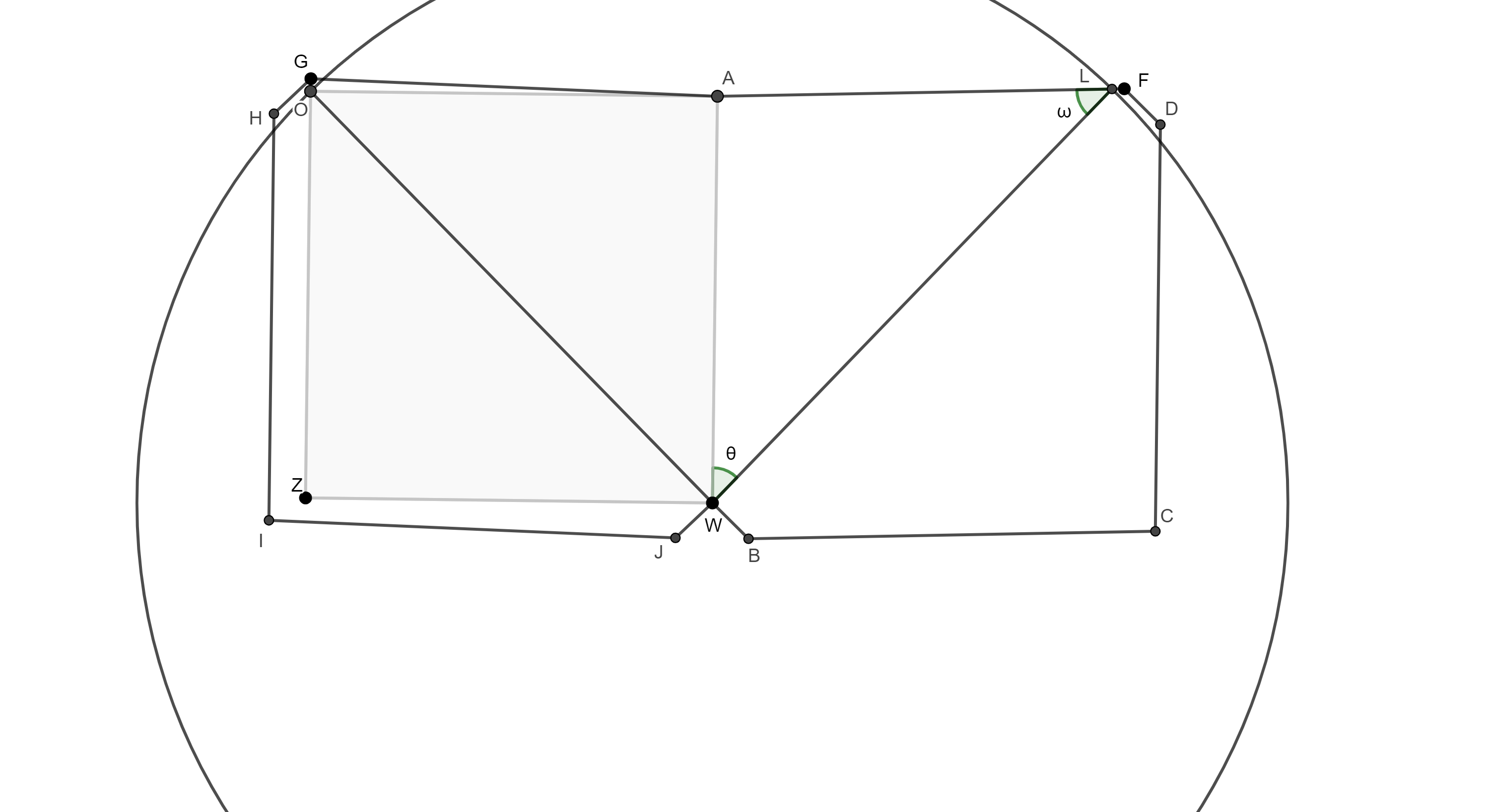}
\caption{Hexagonal lattice and $\zeta(5)$ }
\end{figure}
The next candidate Brillouin zone of the hexagons' conjecture gives a ratio
\begin{equation}
	\frac{AF}{AL}=1.00769169\approx \zeta(7)\rightarrow\frac{AL}{AF}\approx\frac{1}{\zeta(7)}
\end{equation}
where the approximation is $ 99.93\%$.
Finally we have the first Brillouin zone of the square lattice with the ratio

\begin{equation}
	\frac{AF}{AL}=\frac{AL}{AF}=1=\zeta(\infty)
\end{equation}

It is very interesting that the value of the golden ratio and Riemann $\zeta$ functions are connected to each other from the generalized golden ratio formula like:

\begin{equation}
	x^n-x^{n-1}-x^{n-2}-\dots-1=0
\end{equation}
with $n=1,2,3\dots$,
creating a "family" of first Brillouin zones whose construction is based on $\zeta$ functions and the golden ration $\phi$.

For example if $n=1$ then the golden ratio equation becomes:
$x-2=0\rightarrow x=2$ and $x-1=1$ (square lattice).
If $n=2$ then the generalized golden ratio equation becomes:
$x^2-x-1=0\rightarrow x=\frac{1+\sqrt{5}}{2}\rightarrow x-1=\frac{1}{\phi}$ (honeycomb lattice). This motive continues like for $n=3$ and $x-1=\frac{1}{\zeta(3)}$, for $n=5$ we have $x-1=\frac{1}{\zeta(5)}$, etc. These are the values of the ratios
\begin{equation}
	\frac{AL}{AF}=\frac{1}{\zeta(n)},
\end{equation}
$n=3,5,7,\dots$
for all the pertubed lattices coming from the regular hexagonal lattice. 

There is also a similar construction of the first Brillouin zones of the triangular dual lattices that correspond to the $CP^{N-1}$ bosonic model at imaginary chemical potential. The first Brillouin zone of the triangular lattice corresponding to Figure 2 is the one of Figure 5. We see now that there is another thermal window of the fermionization of the model.

\begin{itemize}
	\item \underline{Figure 5}
\end{itemize} 

$AF=AW=\alpha,WL=WO=\sqrt{2}\alpha$,
$\frac{AW}{sin\omega}=\frac{AL}{sin\theta}=\frac{WL}{sin\pi/3}\rightarrow \frac{\alpha}{sin\omega}=\frac{AL}{sin\theta}=\frac{WL}{sin\pi/3}=\frac{2\sqrt6 \alpha}{3}$, so $sin\omega=\frac{\sqrt6}{4}$,
where $\omega$ is the angle between AL and WL and $\theta$ is the angle between AW and WL. 
Also, $sin^2\omega+cos^2\omega=1\rightarrow cos\omega=\frac{\sqrt10}{4}$. But $\theta+\omega=\frac{2\pi}{3}\rightarrow$

$sin\theta=sin(\frac{2\pi}{3}-\omega)\rightarrow sin\theta=sin\frac{2\pi}{3}\cdot cos\omega-cos\frac{2\pi}{3}\cdot sin\omega=\frac{\sqrt6}{8}(\sqrt5 +1).$

So, $AL=sin\theta\cdot\frac{2\sqrt6 \alpha}{3}=\frac{(\sqrt5 +1)\alpha}{2}=\alpha\phi$. At the end we find that
\begin{equation}
	\frac{AF}{AL}=\frac{1}{\phi}\rightarrow\frac{AL}{AF}=\phi
\end{equation}
with the corresponding triangular lattice (Figure 5).

\begin{figure}
\centering
\includegraphics[scale=0.34]{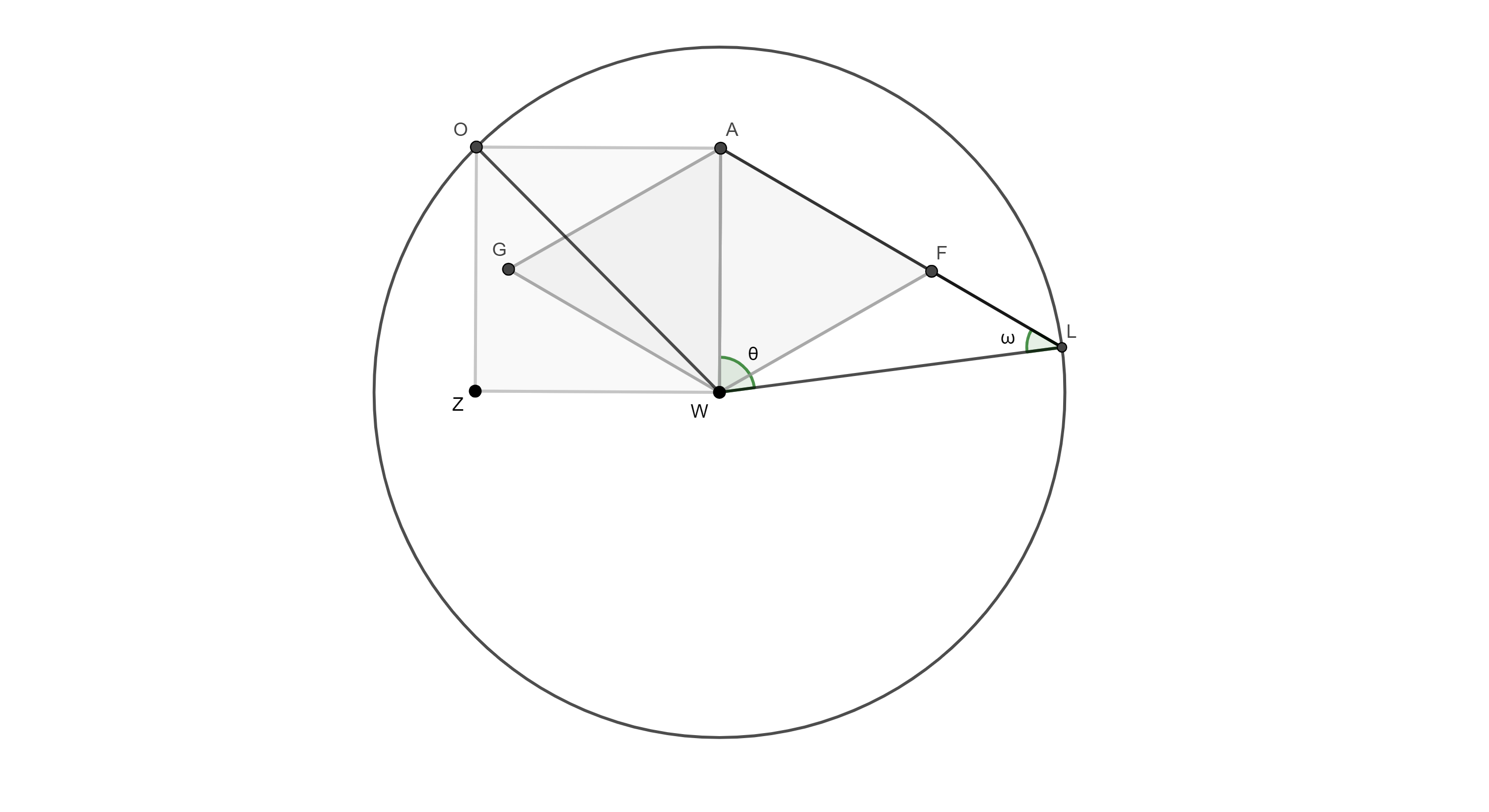}
\caption{Regular triangular lattice and golden ratio $\phi$ }
\end{figure}
The next lattice (Figure 6) is:

\begin{figure}
\centering
\includegraphics[scale=0.34]{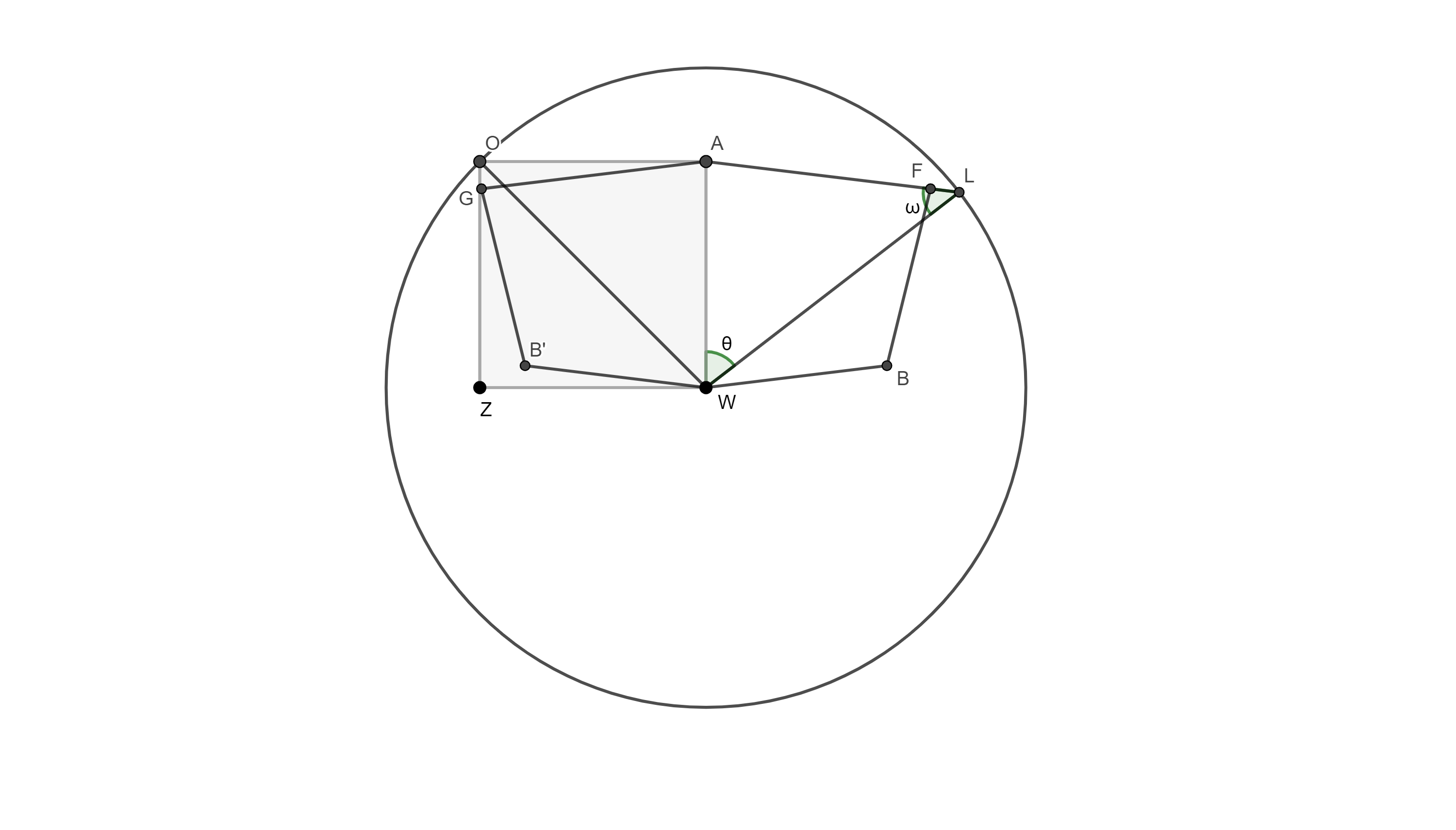}
\caption{Quadrilateral lattice and $\zeta(3)$ }
\end{figure}

$AF=AW=\alpha,WL=WO=\sqrt{2}\alpha$,
$\frac{AW}{sin\omega}=\frac{AL}{sin\theta}=\frac{WL}{sin 6\pi/13}\rightarrow \frac{\alpha}{sin\omega}=\frac{AL}{sin\theta}=\frac{WL}{sin 6\pi/13}=1.4246 \alpha$, so $sin\omega=0.70195$,
where $\omega$ is the angle between AL and WL and $\theta$ is the angle between AW and WL. 
Also, $sin^2\omega+cos^2\omega=1\rightarrow cos\omega=0.712225$. But $\theta+\omega=\frac{7\pi}{13}\rightarrow$

$sin\theta=sin(\frac{7\pi}{13}-\omega)\rightarrow sin\theta=sin\frac{7\pi}{13}\cdot cos\omega-cos\frac{7\pi}{13}\cdot sin\omega=0.791651474.$

So, $AL=sin\theta\cdot 1.4246\alpha=1.12778669\alpha$. At the end we find that
\begin{equation}
	\frac{AF}{AL}=\frac{1}{1.12776}\rightarrow\frac{AL}{AF}=1.12778669\approx \zeta(3)
\end{equation}
which is exactly the inverse of the result of the corresponding transformed hexagonal lattice (Figure 3). 

This motive continuous but now we are inetrested in $\frac{1}{x-1}$ solutions of equation $x^n-x^{n-1}-x^{n-2}-\dots-1=0$. These are $\phi, \zeta(3), \zeta(5), \dots, \zeta(\infty)$.

{\centering
	\begin{tabular}{|c|c|c|c|c|}
		\hline 
		\multicolumn{5}{|c|}{\textbf{Table 2. The interior angles of quadrilaterals conjecture-Euclidean space}} \\ 
		\hline 
		Dimensions & Angle $1$ & Angle $2$ & Angle $3$ & Angle $4$ \\ 
		\hline 
		5 & $6\pi/13$ & $6\pi/13$ & $6\pi/13$ & $8\pi/13$ \\ 
		\hline 
		7 & $25\pi/51$ & $25\pi/51$ & $25\pi/51$ & $27\pi/51$ \\ 
		\hline 
		9 & $102\pi/205$ & $102\pi/205$ & $102\pi/205$ & $104\pi/205$ \\ 
		\hline 
		... & ... & ... & ... & ...  \\ 
		\hline 
		$\infty$ & $\pi/2$ & $\pi/2$ & $\pi/2$ & $\pi/2$ \\ 
		\hline 
	\end{tabular}\par}

Since I show that the $Gross-Neveu$ and $CP^{N-1}$ models at imaginary chemical potentials are duals for a specific shift of the chemical potential we may say that fermions and bosons "see" the dual lattices as they move on the surfaces of them.

Tables 1 and 2 seems to be clearly in agreement with equation  (8), where at infinite dimensions the total transfer momentum of the fundamental particles is equal to the overall transfer momentum of the modified Bragg Law, apart from a factor arising from the compact dimension on the thermal circle that disappears at infinite dimensions. Although I go physically to higher dimensions, it can be reverted to a study of 2 + 1 dimensions by scattering into hypothetical crystals, which (with very good approximation) give us the changes of the momentum of the scattered particles. We also have to mention that we will have the same results if we use the angles $6\pi/13$, $25\pi/51$, $102\pi/205$.. with the corresponding quadrilaterals of our conjecture where the first Brillouin zone of the family is the one of the triangular lattice (dual lattice) and the construction is made of tetrahedra instead of hexagons. Like before the "infinite dimensional theory" construction turns to a square/rectangle tiling of the original Bragg Law.

It is also interesting that there are two kind of functions that become from the ratio $\frac{AL}{AF}$ and take values 
\begin{align}
	F(x)=\frac{1}{K}, G(x)=K 
\end{align}
where $K=1,\phi,\zeta(3),\zeta(5)\dots$. The product of these functions 
\begin{equation}
	F(x)G(x)
\end{equation}
is a normalization value that equals to $1$ for the $3d$ case and also for the other cases. The ratios are almost $\zeta(n)$ or $\frac{1}{\zeta(n)}$ and this may come from the fact that theories in higher dimensions are non-renormalizable and the lattices that arise from the conjectures do not have normal polygons. I conjugate that this product gives the duality picture of the hexagonal and triangular lattices. It is like the duality image of a fermionic and bosonic theory where their partition functions are inverse one of the other.

One may see the transformation of a triangular lattice to a hexagonal one and compare with the phase transitions of the Gross-Neveu model at imaginary chemical potential. The picture is like the one that follows:

''Studying the correspondence of thermal windows with the transformation of a hexagonal to triangular lattice we have fermions at a high temperature and freeze them until we reach a critical temperature where we go to boson condensates. The thermal window closes at a lower temperature and we have fermions again. Respectively in some hexagonal materials we have phase transitions with a broken symmetry where triangular materials are arising (like the FCC to BCC transformations).''

\section{Generalised thermal windows and lattice points}

Let's see in detail Figure 2. We have a circle with radius $\sqrt2a$ that intersects segment $AF$ at $L$ point. Interestingly there is an equivalent picture of the thermal windows in the Gross-Neveu model at finite T and imaginary chemical potential. In the Gross-Neveu model and its change in the statistics we have temperature (the chemical potential) that "creates" the thermal windows, where inside them we have a chiral symmetry breaking in order to make some fermion condensates. In our conjecture the thermal windows have the equivalent picture of the region inside the hexagon. As we will see later the borders of the thermal windows coincides with the positions of the lattice points of the hexagon. At these points, $D_2(-z*)$ has its maximum value. The lattice point D is the equivalent point of the full bosonization in the Gross-Neveu model. The points L and F are coming closer and at $d\rightarrow\infty$ they coincide.  One may say it seems that at infinite dimensions symmetry is always unbroken and we return to a fermionic theory.
	
	The generalised thermal windows are:\\
	
	{\centering
		\begin{tabular}{|c|c|c|}
			\hline 
			\multicolumn{3}{|c|}{\textbf{Table 3. Generalized Thermal Windows for the $GN$ model}} \\ 
			\hline 
			Dimensions & Closing T & Opening T \\ 
                        \hline
                        3& $\frac{3\alpha_*}{2\pi}\frac{1}{2+3k}$&$\frac{3\alpha_*}{2\pi}\frac{1}{1+3k}$\\
			\hline 
			5 & $\frac{13\alpha_*}{7\pi}\frac{1}{\frac{19}{7}+\frac{26k}{7}}$ & $\frac{13\alpha_*}{7\pi}\frac{1}{1+\frac{26k}{7}}$ \\ 
			\hline 
			7 & $\frac{51\alpha_*}{26\pi}\frac{1}{\frac{76}{26}+\frac{102k}{26}}$ & $\frac{51\alpha_*}{26\pi}\frac{1}{1+\frac{102k}{26}}$ \\ 
			\hline 
			9 & $\frac{205\alpha_*}{103\pi}\frac{1}{\frac{307}{103}+\frac{410k}{103}}$ & $\frac{205\alpha_*}{103\pi}\frac{1}{1+\frac{410k}{103}}$ \\ 
			\hline 
			11 & $\frac{819\alpha_*}{410\pi}\frac{1}{\frac{1228}{410\pi}+\frac{1638k}{410}}$ & $\frac{819\alpha_*}{410\pi}\frac{1}{1+\frac{1638k}{410}}$ \\ 
			\hline 
		\end{tabular}\par} 
	where $k =0 ,1 ,2 ,\ldots$.

Let's focus on the case $k=0$ of the $3d$ theory thermal window. The borders for $\alpha_*$ are $2\pi T/3$ and $4\pi T/3$. These are the points where the $D_2(-z*)$ takes its maximum value (imaginary part) on the unit circle. On the unit circle $D_2(-z*)=Cl_2(\pi-\beta\alpha_*)$. The generalised thermal windows for the $CP^{N-1}$ model are:\\

 {\centering
		\begin{tabular}{|c|c|c|}
			\hline 
			\multicolumn{3}{|c|}{\textbf{Table 4. GeneraliSed Thermal Windows for the $CP^{N-1}$ model}} \\ 
			\hline 
			Dimensions & Closing T & Opening T \\ 
                        \hline
                        3& $\frac{3\alpha_*}{2\pi}\frac{1}{\frac{5}{2}+3k}$&$\frac{3\alpha_*}{2\pi}\frac{1}{\frac{1}{2}+3k}$\\
			\hline 
			5 & $\frac{13\alpha_*}{6\pi}\frac{1}{\frac{20}{6}+\frac{26k}{6}}$ & $\frac{13\alpha_*}{6\pi}\frac{1}{1+\frac{26k}{6}}$ \\ 
			\hline 
			7 & $\frac{51\alpha_*}{25\pi}\frac{1}{\frac{77}{25}+\frac{102k}{25}}$ & $\frac{51\alpha_*}{25\pi}\frac{1}{1+\frac{102k}{25}}$ \\ 
			\hline 
			9 & $\frac{205\alpha_*}{102\pi}\frac{1}{\frac{308}{102}+\frac{410k}{102}}$ & $\frac{205\alpha_*}{102\pi}\frac{1}{1+\frac{410k}{102}}$ \\ 
			\hline 
			11 & $\frac{819\alpha_*}{409\pi}\frac{1}{\frac{1229}{409\pi}+\frac{1638k}{409}}$ & $\frac{819\alpha_*}{409\pi}\frac{1}{1+\frac{1638k}{409}}$ \\ 
			\hline 
		\end{tabular}\par} 
	where $k =0 ,1 ,2 ,\ldots$.

Let's focus again on the case $k=0$ of the $3d$ theory thermal window. The borders for $\alpha_*$ are $\pi T/3$ and $5\pi T/3$. These are the points where the $D_2(z*)$ takes its maximum value (imaginary part) on the unit circle. On the unit circle $D_2(z*)=Cl_2(\beta\alpha_*)$. The most interesting part is to put all these points on the unit circle which is circumscribed about a regular hexagon and a triangle. We see that at $\alpha_*=\pi T$, $D_2(z*)$ takes its maximum value (real part) and at $\alpha_*=0$, $D_2(-z*)$ takes its maximum value (real part). So the lattice points of the "supersymmetric" model are lying on the unit circle at $\alpha=0, \pi/3, 2\pi/3, \pi, 4\pi/3, 5\pi/3$. The "supersymmetric" model when the regular hexagon turns to square has a duality picture between fermions and bosons. If we see it from the boson sight of view we have bosonic matter at the right side of the unit circle. At the left side bosonic matter turns to fermionic. From the fermion sight of view we have fermionic matter at the right side of the unit circle and at the left side fermionic matter turns to bosonic. The overall model has mirror particles that we can not distinguish them.

\begin{figure}
               \centering
                \includegraphics[scale=0.44]{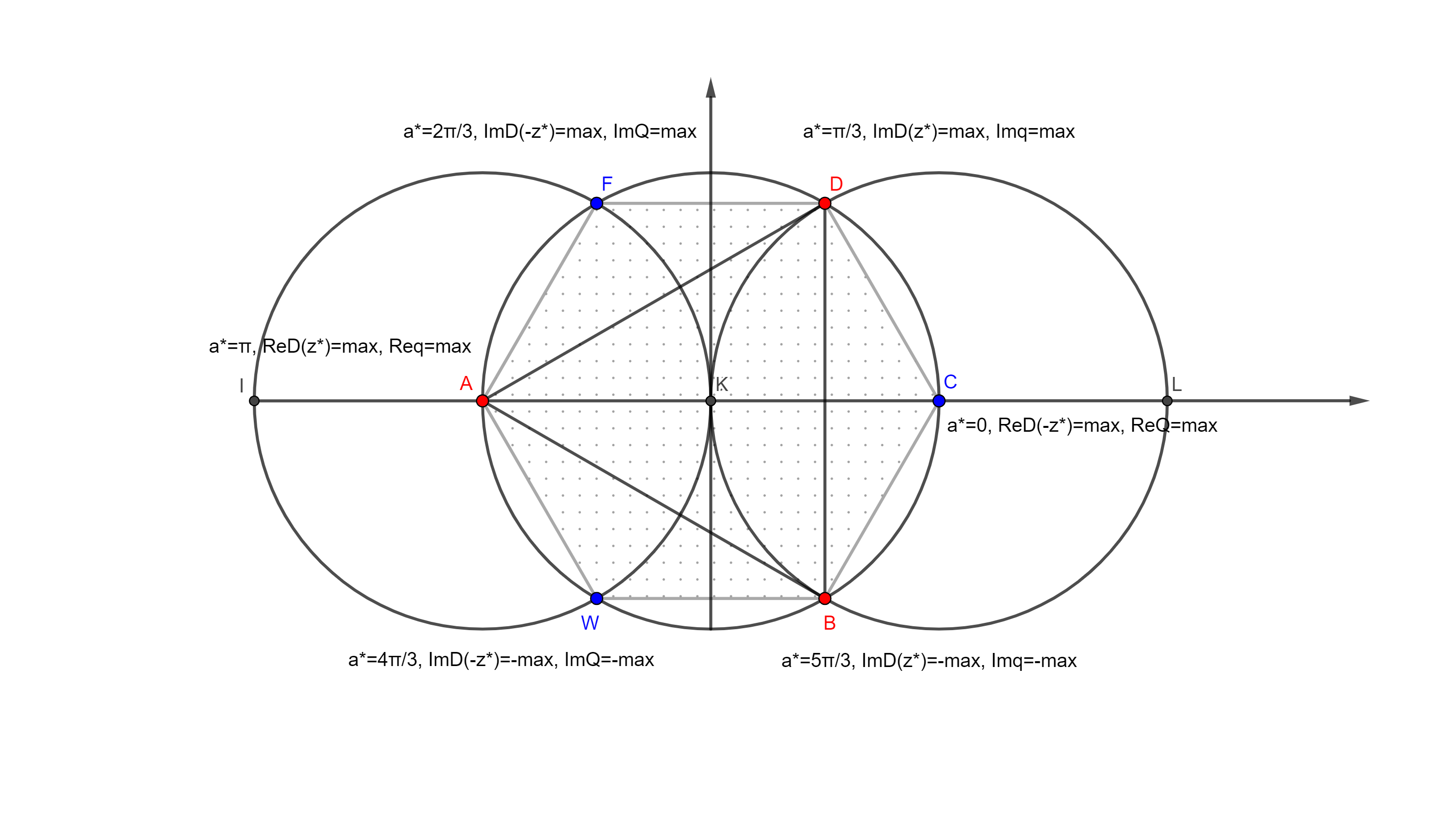}
                 \caption{GN and CP models on lattices}
\end{figure}

\begin{figure}
               \centering
                \includegraphics[scale=0.68]{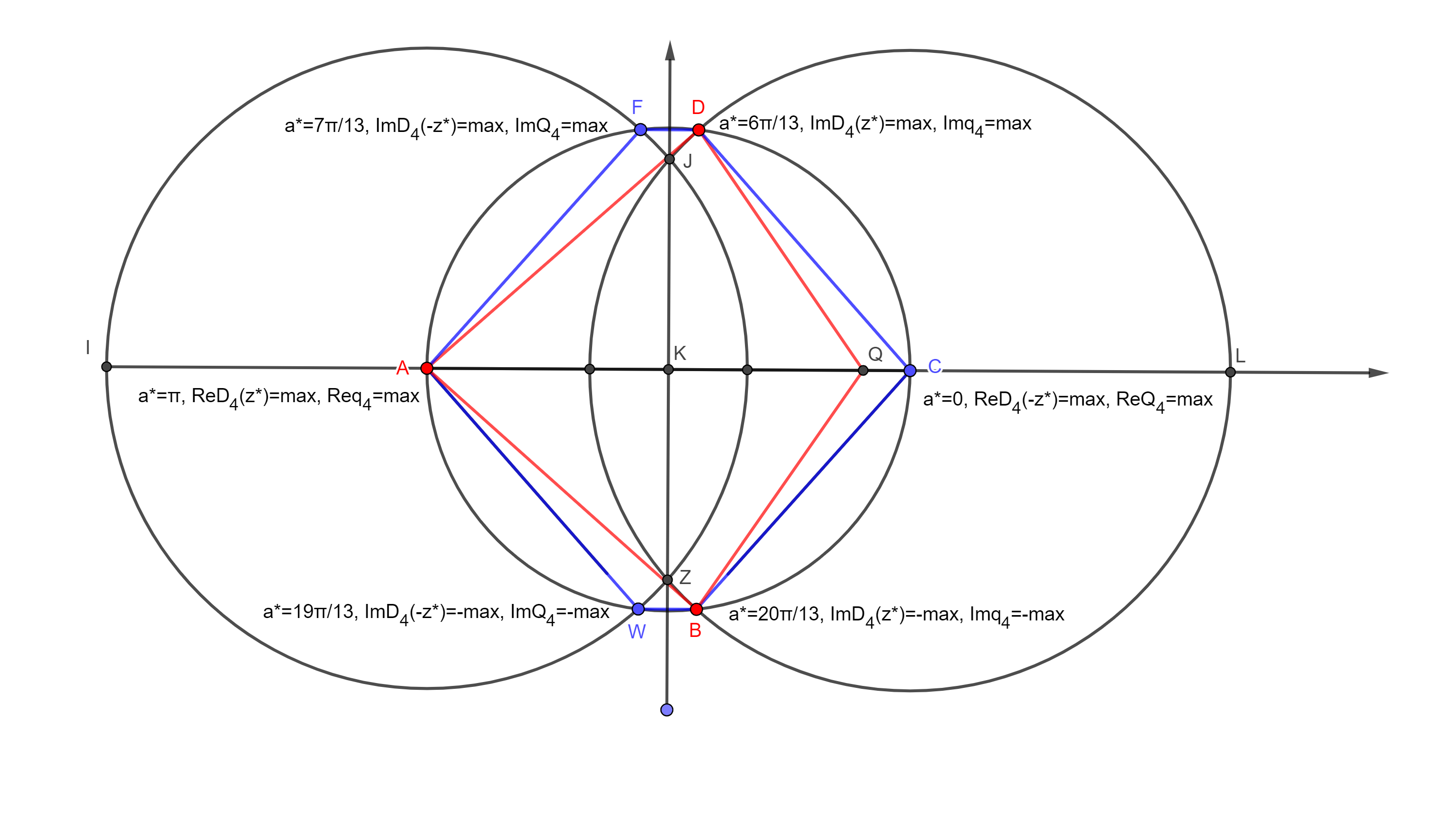}
                 \caption{GN and CP models on perturbed lattices}
\end{figure}

\begin{figure}
               \centering
                \includegraphics[scale=0.44]{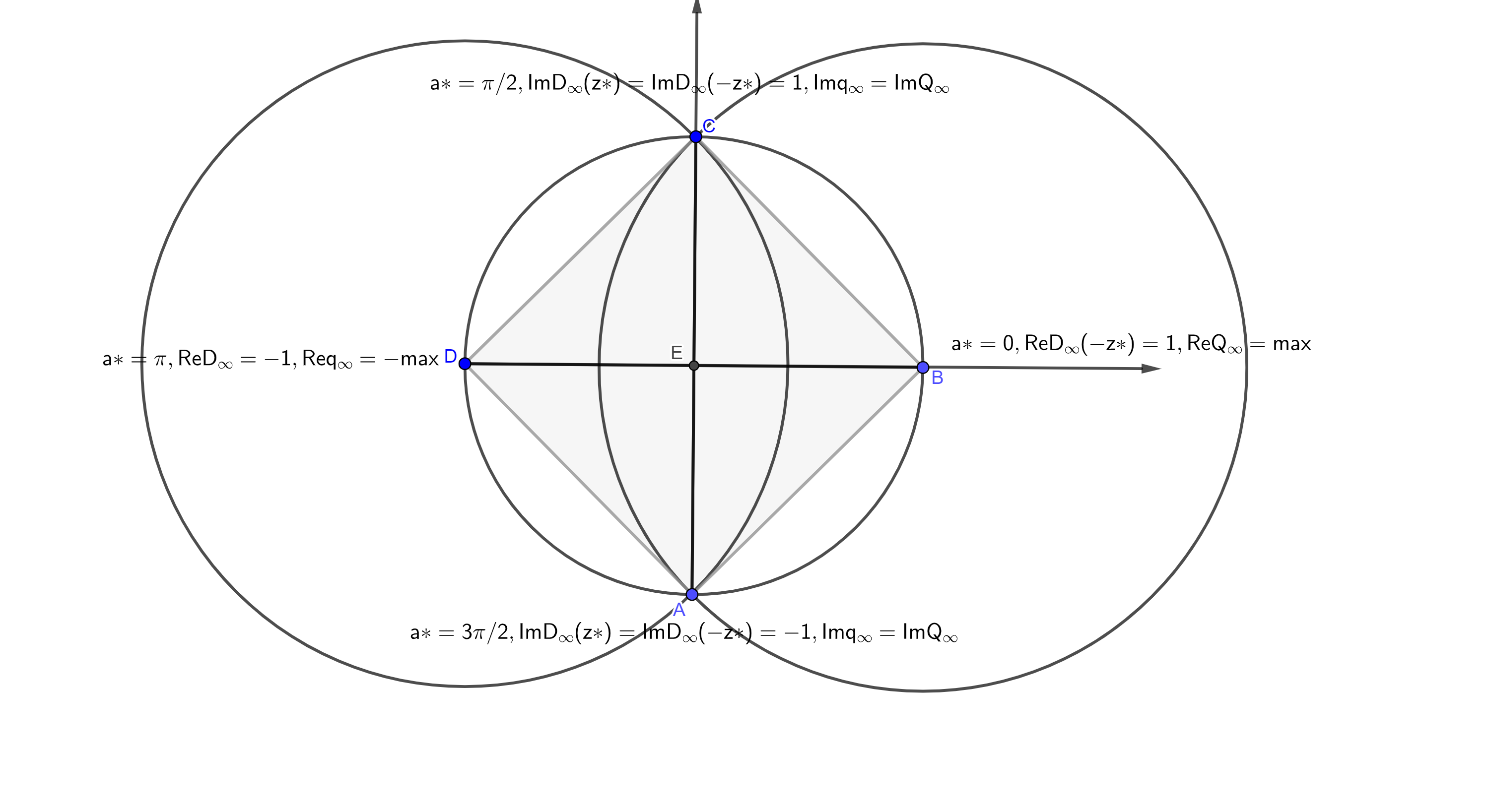}
                 \caption{GN and CP models on square lattice}
\end{figure}

\section{Fermion and boson partition functions as Bloch waves}
	
	The modification of equation (2) is the transfer momentum for arbitrary odd dimensions $m$ which is of the form:
\begin{equation}
		\bar Q_{total}=4\pi N Cl_{m}(\theta)=4\pi N (sin\theta +\sum_{k=2}^{\infty}\frac{sink\theta}{k^m})
	\end{equation}
In my previous work I noticed that the fermionic and bosonic gap equations of the fermionic and bosonic models at finite T and imaginary chemical potential are expressed in linear combinations of Bloch-Wigner-Ramakrishnan functions, analysed by Zagier. The $D_d(z)$ are real valued complex functions, with their best known representative the $D(z)= D_2(z)$ function which gives the volume of ideal tetrahedra in Euclidean Hyperbolic space $\mathbb H_3$ whose four vertices lie on $\partial\mathbb H_3$ at the points $0,1,\infty, z$. These tetrahedra are the building blocks for general hyperbolic manifolds - the volume of the latter arises as the sum of ideal tetrahedra after a suitable triangulation (It would be interesting to examine the triangulation of $5d$-manifolds \cite{Fluder}) .	
	
	The difference with Bragg's law that gives the transfer momentum from a beam of incident particles to the target lattice is the second term of the sum. This term at the limit where $d$ goes to infinity is zero and we end up with Bragg's well-known law of scattering. The parameter $d$ somehow plays the role of reducing the effect of the dimensional reduction from the compactification of the time-dimension, returning things to their original condition.This model is a transformation path from a hexagonal and triangular lattice to a square lattice where we have the extreme values of quasimomentum $\bar{Q}_{total}$ for specific angles of beam-scattering with an a excellent approximation.
	
	From the moment when $Cl_m(\theta)$ function is periodic, the last relationship may come from a periodic crystal. If we think that the maximization of the function is done for specific values of the angle like $2\pi/3$ or equally $\pi/3$, a filling of the crystal plane based on figures 1-5 can describe the phenomenon. Seeking the proper filling of the plane with atoms of a hypothetical crystal, the so-called tiling, we may use as a candidate the hexagonal and triangular lattices. This choice is based on the fact that when particles fall with $2\pi/3$ or $\pi/3$ angles at any direction defined by the atoms at the vertices of the regular hexagon or triangle, their overall transfer momentum is maximized. The question now is how we express the wave function of a particle that travels inside a periodic potential of a crystal structure. The appropriate expression is a Bloch-wave of the form:

\begin{equation}
\Psi_{n\overrightarrow {k}}=e^{i\overrightarrow {k}\overrightarrow {r}}U_{n\overrightarrow {r}}
\end{equation}
where $U$ function has the periodicity of the lattice. On the other hand consider a system at finite temperature $T=1/\beta$ with a global $U(1)$ charge operator $\hat{Q}$ with the Fourier transform of the grand canonical partition function:
\begin{equation}
Z_c(\beta,Q)=\int_{0}^{2\pi}\!\frac{d\theta}{2\pi}\,e^{i\theta Q}\,\rm Tr\left[e^{-\beta\hat{H}-i\theta\hat{Q}}\right]=\int_{0}^{2\pi}\!\frac{d\theta}{2\pi}\,e^{i\theta Q}\,Z_{gc}(\beta,\mu=-i\theta/\beta),
\end{equation}
These are similar expressions where $U$ functions are the odd index $D$ functions and the exponential of $Q$ part (imaginary part of $Q$) is the even index $D$ functions. There is a possible expression of scattering in one of the above lattice points of wave functions that look like Bloch waves which give a general expression of the transferred momentum to the lattice as a function of the Clausen functions. At the large limit the expression ends in Bragg law. These Bloch waves consist of two parts. One with odd $D (z)$ and one with even $D (z)$. Functions with $D_{odd}$ are orthogonal to the unit and do not contribute to the transferred momentum. The contribution is made only by $D_{even}$.  The changes of $k$ vector (in our case Q) is like one adds a reciprocal lattice vector to $k$ where all the resulting waves are equal and they constitute a Bloch state.  The extra dimensions of the continuous theory is like we add an extra vector to $k$ that express the possible conditions that a Bloch wave may be found. If we limit the phenomenon in $1$st Brillouin zone all the new $k$ are different and they are like the charges in odd dimensions. So, as the dimension increases we have a compression of Bloch states in the $1$st Brillouin zone. The function $U(r)$ has the periodicity of the lattice where the Bloch wave travels. In continuous theory this periodicity comes from the chemical potential. These functions $U_{\left(n\overrightarrow{r}\right)}, U_{\left(n'\overrightarrow{r'}\right)}$ are normalized to unity so the only contribution to the product $\Psi_{\left(n\overrightarrow{r}\right)} \Psi_{\left(n'\overrightarrow{r'}\right)}$ comes from the $k$ exponential. 

	\begin{figure}
		\centering
		\includegraphics[scale=0.42]{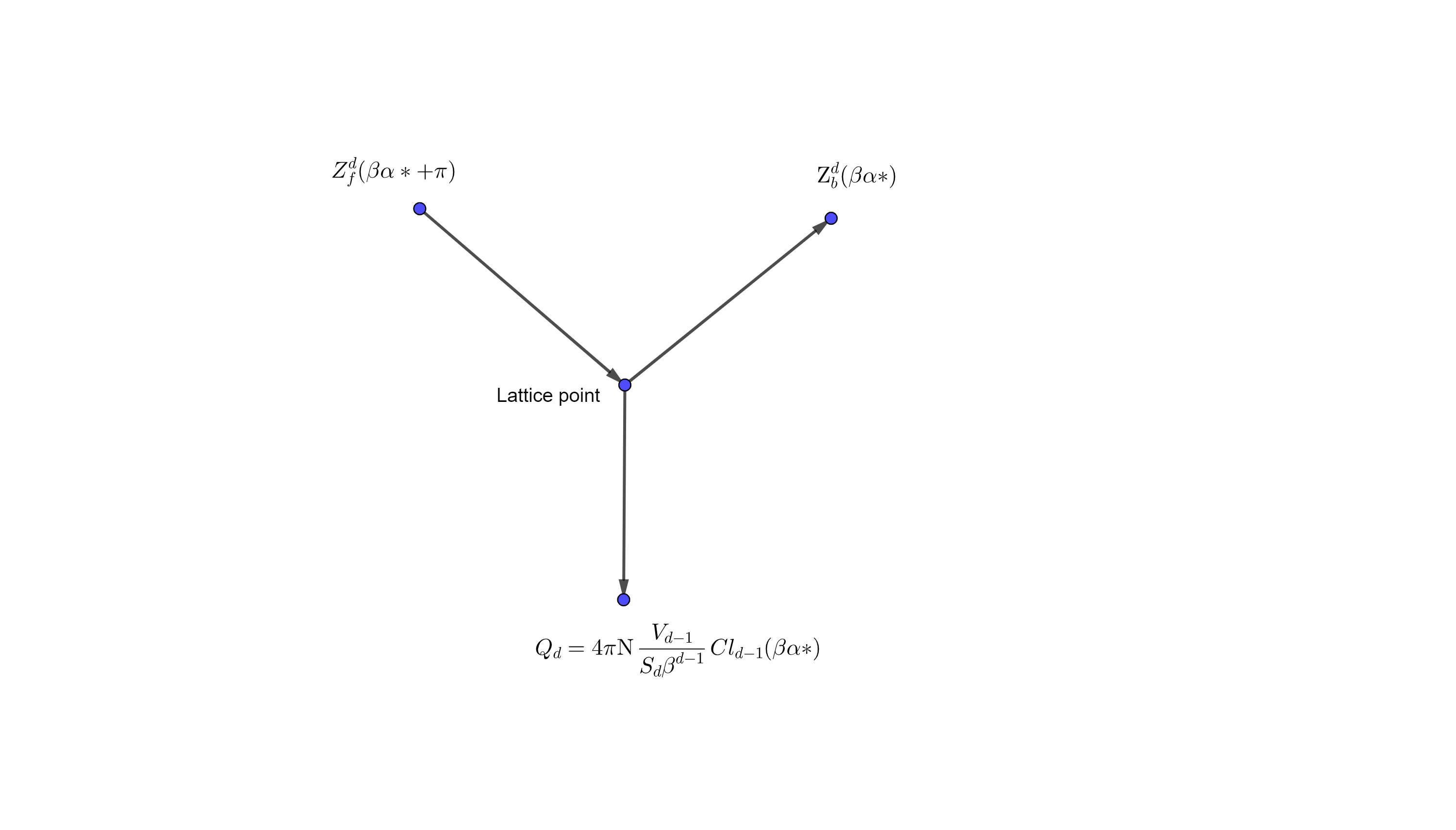}
		\caption{Bloch waves on lattice point}
	\end{figure} 
	
	Consider that on every hexagon's vertex there is an atom of the crystal structure. The conjectures of my analysis are coming to the final point. When we have a charged fermionic and bosonic model, where the fundamental degrees of freedom are the basis of our physics (which means that we have no condensates), the total generating functional for the "supersymmetric" model is giving at $\sigma=0$ the overall transfer momentum of $N$ particles that are scattered from an hypothetic crystal with specific construction based on generalized golden ratio generating formula. The imaginary chemical potential and its changes plays the role of the various guided angles of the particle beams on the lattice. This is not only an interesting feature of a "supersymmetric" model of charged particles that scattered from a lattice but also a theoretical observation of the peaks of scattering for odd dimensional physics that we will guide us for future works. These peaks correspond to the well-known $Bragg-Peaks$ \cite {Hickey, Jin, Liu} on a Bragg curve which plots the energy loss of ionizing radiation during its travel through matter. These peaks occurs because the interaction cross section increases as the charged particle's energy/momentum decreases.

	Since I study particle models, with the presence of an imaginary chemical potential, I am then be able to give a more general effect on scattering of particles of specific lattices. The very interesting result I export gives a fairly accurate picture of these scatterings and allows the indirect analysis of such a phenomenon into microscopic models and up to infinite theoretical dimensions.
Let's see now the overall result for the fermion and boson partition functions in $5d$ and arbitrary odd $d$.
\begin{equation}
Z_{tot}^{(5)}(\beta\alpha_*)\equiv Z_f^{(5)}(\beta\alpha_*+\pi)\left[Z_{b}^{(5)}(\beta\alpha_*)\right]^{\frac{{\rm Tr}{\mathbb I}_{4}}{2}}=e^{-iNV_4\frac{\pi}{6}q_5}\,.
\end{equation}
To generalize for all $d$ we note firstly that the exponent ${\rm Tr}{\mathbb I}_{d-1}/2$  of the bosonic partition function is a manifestation of the usual fact that as we go up in dimension a single Dirac fermion "weights" more and more bosonic degrees of freedom. Hence, if we choose a model with a supersymmetric matter content,\footnote{We do not want imply that there is some sort of  supersymmetry here, as this would require many more additional d.o.f.} such that the number of bosons is $2^{\frac{d-1}{2}}$, then we could get rid of that $d$-dependent exponent. At the end we can give the formula for the partition function duality between fermionic and bosonic theories, with supersymmetric matter content, at $\sigma_*=0$ and for general $d$ as
\begin{equation}
\label{susyd}
Z^{(d)}_{"SUSY"}(\beta\alpha_*)\Bigl|_{\sigma_*=0}\equiv\left[Z_f^{(d)}(\beta\alpha_*+\pi)Z_{b}^{(d)}(\beta\alpha_*)\right]\Bigl|_{\sigma_*=0}=e^{4\pi N\frac{V_{d-1}}{S_d \beta^{d-1}}Cl_{d-1}(\beta\alpha_*)}
\end{equation}
The result suggests the existence of a non trivial large-$d$ limit of the fermion-boson duality. Namely, if we take the zero temperature or decompactification limit $\beta\rightarrow\infty$ the ratio $V_{d-1}/S_{d}\beta^{d-1}\rightarrow 1$ (i.e. we can think of $V_{d-1}$ as the surface of a very large sphere). Moreover, the Clausen's functions with even index, such as those that appeared before, have a well defined $d\rightarrow\infty$ limit which is simply $\sin(\beta\alpha_*)$. Therefore we can write
\begin{equation}
\label{limsusy}
\lim_{d\rightarrow\infty} Z^{(d)}_{"SUSY"}(\beta\alpha_*)\Bigl|_{\sigma_*=0}=e^{4\pi N\sin(\beta\alpha_*)}\,
\end{equation}
where the exponential is the well known transfer momentum of the Bragg Law (\ref{Bragg}) .

% Numbered equation
%\begin{eqnarray}
%\end{eqnarray}

% Table
%\begin{table}[!htbp]
% Use starred version (\begin{table*} \end{table*}) for two column spanned table
%\caption{\label{tbl1}Caption of the table.}
%\centering 
%\def\arraystretch{1.15} % To increase the row spacing
%\begin{tabular}{ll}
%\hline
%One & Two \\
%\hline
%Three & Four\\
%\hline
%\end{tabular} 
%\end{table}

\section{Deep inside the strong coupling regime}
 To calculate the condensate gap-equation of the $U(N)$ Gross-Neveu model in arbitrary odd dimensions and in the presence of imaginary chemical potential $\mu=-i\alpha$ I use the Euclidean action .
 \begin{equation}
 	S_{GN} = -\int_0^\beta \!\!\!dx^0\int \!\!d^{d-1}\bar{x} \left[\bar{\psi }^{a}(\slash\!\!\!\partial  -i\gamma_0\alpha)\psi ^{a}+\frac{G_d}{2({\rm Tr}\mathbb I_{d-1})N}\left (\bar{\psi }^{a}\psi ^{a}\right )^{2} +i\alpha NQ_d\right]\,,\,\,\,a=1,2,..N.
 \end{equation}
 
 where $Q_d$ is the eigenvalue of the $N$-normalized fermion number density operator $\hat{Q}_d=\psi^{a\dagger}\psi^a/N$ in $d$ odd dimensions and it comes from the Lagrangian of the model that possesses a $U(1)$ global symmetry. Introducing an auxiliary scalar field $\sigma$ the canonical partition function is given by
 \begin{equation}
 	S_{f,eff}=iQ_d\int_0^\beta \!\!\!dx^0\!\!\int\!\!d^{d-1}\bar{x}\,\alpha-\frac{{\rm Tr}\mathbb I_{d-1}}{2G_d}\int_0^\beta \!\!\!dx^0\int d^{d-1}\bar{x}\sigma^2+\rm Tr\ln(\slash\!\!\!\partial-i\gamma_0\alpha+\sigma)_\beta
 \end{equation}
 
 To evaluate the condensate gap equation we look for constant saddle points $\alpha_*$ and $\sigma_*$. At large-$N$ we have the gap equation
 \begin{equation}
 	\frac{\partial}{\partial\sigma}S_{f,eff}\Biggl|_{\sigma_*,\alpha_*}\!\!\!\!=0\,\,\,\Rightarrow\,\,\, -\frac{\sigma_*}{G_d}+\frac{\sigma_*}{\beta}\sum_{n=-\infty}^\infty\int^\Lambda\!\!\frac{d^{d-1} \bar{p}}{(2\pi)^{d-1}}\frac{1}{\bar{p}^2+(\omega_n-\alpha_*)^2+\sigma_*^2}=0
 \end{equation}
 where  the fermionic Matsubara sums are over the discrete frequencies $\omega_n=(2n+1)\pi/\beta$. The divergent integrals are regulated by the cutoff $\Lambda$. 
 The main issue with the GN model in $d>3$ is that the gap equation has  a finite number of higher order divergent terms as $\Lambda\rightarrow\infty$, which cannot be simply taken care of by the adjustment/renormalization of the single coupling $G_d$.To avoid these divergent parts I have to try a kind of lattice regularization which is a way to replace the space-time continuum of the particle movements with a discrete set of lattice points where $D_{odd}(-z*)=0$ at the highest odd dimension and at the critical point of the corresponding field theory. The lower dimensions have their own gap equations inside the gap equation of the $d$ model and I regularize the theory by replacing the cut-offs with a mass scale that separates the weak from the strong coupling regime. Interestingly the values of the mass scales have specific relationship with the places of the lattice points of the discrete conjectures. This technique provides somehow a kind of definition of the theory which has unspecified quantities in the continuum space-time regime.

 \begin{itemize}
 	\item \underline{3d gap equation}
 \end{itemize}
 
 The gap equation in 3 dimensions of the Gross-Neveu model at imaginary chemical potential at the critical point $M_3=0$ and for $\alpha_*=\pi$ turns to (assuming dimensionless equations):
 
 $\sigma_* D_1(e^{-\sigma_*})=0\rightarrow D_1(e^{-\sigma_*})=0\rightarrow \ln(1-e^{-\sigma_*})-\ln\frac{|e^{-\sigma_*}|}{2}=0\rightarrow \sigma_*=2\ln\phi$ \cite {Filothodoros:2016txa,Sachdev:1993pr}.

 \begin{itemize}
 	\item \underline{5d gap equation}
 \end{itemize}
 
 The gap equation in 5 dimensions of the Gross-Neveu model at imaginary chemical potential at the critical point and for $\alpha_*=\pi$ turns to (assuming dimensionless equations):
 
 $\sigma_*[-M_5-D_3(-z_*)-\frac{\ln^2|z_*|}{2}(D_1(-z_*)-\frac{2\gamma}{3\pi})]=0$, where $\gamma=\Lambda$ the cut-off. If we set $\gamma=M_3$ and put the $5d$ theory at the critical point $M_5=0$ and assume that $D_1(-z_*)-\frac{2M_3}{3\pi}=0$, we find that $M_3=4,12525$ and from the gap equation $D_3(-z_*)=0$ we find that $\sigma_*=2,03185$. It is interesting that $M_3$ is a linear combination of $\zeta$ functions like $M_3=4,12525=3\zeta(3)+\frac{\zeta(5)}{2}$.

 \begin{itemize}
 	\item \underline{7d gap equation}
 \end{itemize}
 
 The gap equation in 7 dimensions of the Gross-Neveu model at imaginary chemical potential at the critical point and for $\alpha_*=\pi$ turns to (assuming dimensionless equations):
 
 $\sigma_*[M_7+D_5(-z_*)+\frac{\ln^2|z_*|}{6}(D_3(-z_*)+\frac{\gamma_1^3}{45\pi})+\frac{\ln^4|z_*|}{24}(D_1(-z_*)-\frac{4\gamma_2}{15\pi})]=0$, where $\gamma=\Lambda$ the cut-off. If we set $\gamma_2=M_3$ and $\gamma_1=M_5$ and put the $7d$ theory at the critical point $M_7=0$, we find that $M_5=6,040316$ and $M_3=16,36421$. Also from the gap equation $D_5(-z_*)=0$ we find that $\sigma_*=2,89218$. It is also interesting that $M_3$ and $M_5$ are linear combination of $\zeta$ functions like:
 $M_5=6,040316=3\zeta(3)+\frac{18\zeta(5)}{5}-\frac{\pi^4}{75}$ and $M_3=16,36421=3\zeta(3)+5\zeta(5)+\frac{7\pi^4}{90}$.
 
The higher dimensions gap equations possibly contain the lower dimension equations giving us an idea of what happens to the strong coupling in lower dimensions when in the upper dimension we are at the critical point $M_{odd}=0$ (the lower dimensions seem to sink even deeper in the strong coupling regime). The projections of the lattice points onto the real axis create line segments, the ratios of which can show us how deep into the strong coupling regime is the $3$-dimensional theory when the higher-dimensional theory is at the critical point $M_d=0$.
The corresponding picture from the hexagon conjecture is:

{\centering
	\begin{tabular}{|c|c|c|c|c|}
		\hline 
		\multicolumn{5}{|c|}{\textbf{Table 5. Inside the strong coupling regime}} \\ 
		\hline 
		Dimensions & $M_3$ & $M_5$ & $M_7$ & $\sigma_*$ \\ 
		\hline 
		3 & $0$ & $-$ & $-$ & $2log\phi$ \\ 
		\hline 
		5 & $4.1252$ & $0$ & $-$ & $2.03185$ \\ 
		\hline 
		7 & $16.36421$ & $6.040316$ & $0$ & $2.89218$ \\ 
		\hline
		$\infty$ & $\infty$ &  $....$ & $....$ & $\infty$ \\
		\hline 
	\end{tabular}\par}
	
where $\phi=1.618$ is the golden ratio.	
It is interesting to see the ratios of figures 11 and 12. The values are exactly the mass scale $M_3$ of the $3d$ theory when we are deeper inside the strong coupling regime.At $d=\infty$ we have the mass scale goes to $\infty$ as well (figure 13). 
The overall picture of the Hexagon conjecture at $\alpha*=\frac{7\pi}{13}$, $\alpha*=\frac{26\pi}{51}$ and $\alpha*=\frac{103\pi}{205}$ is the one that follows:

\begin{figure}
		\centering
		\includegraphics[scale=0.66]{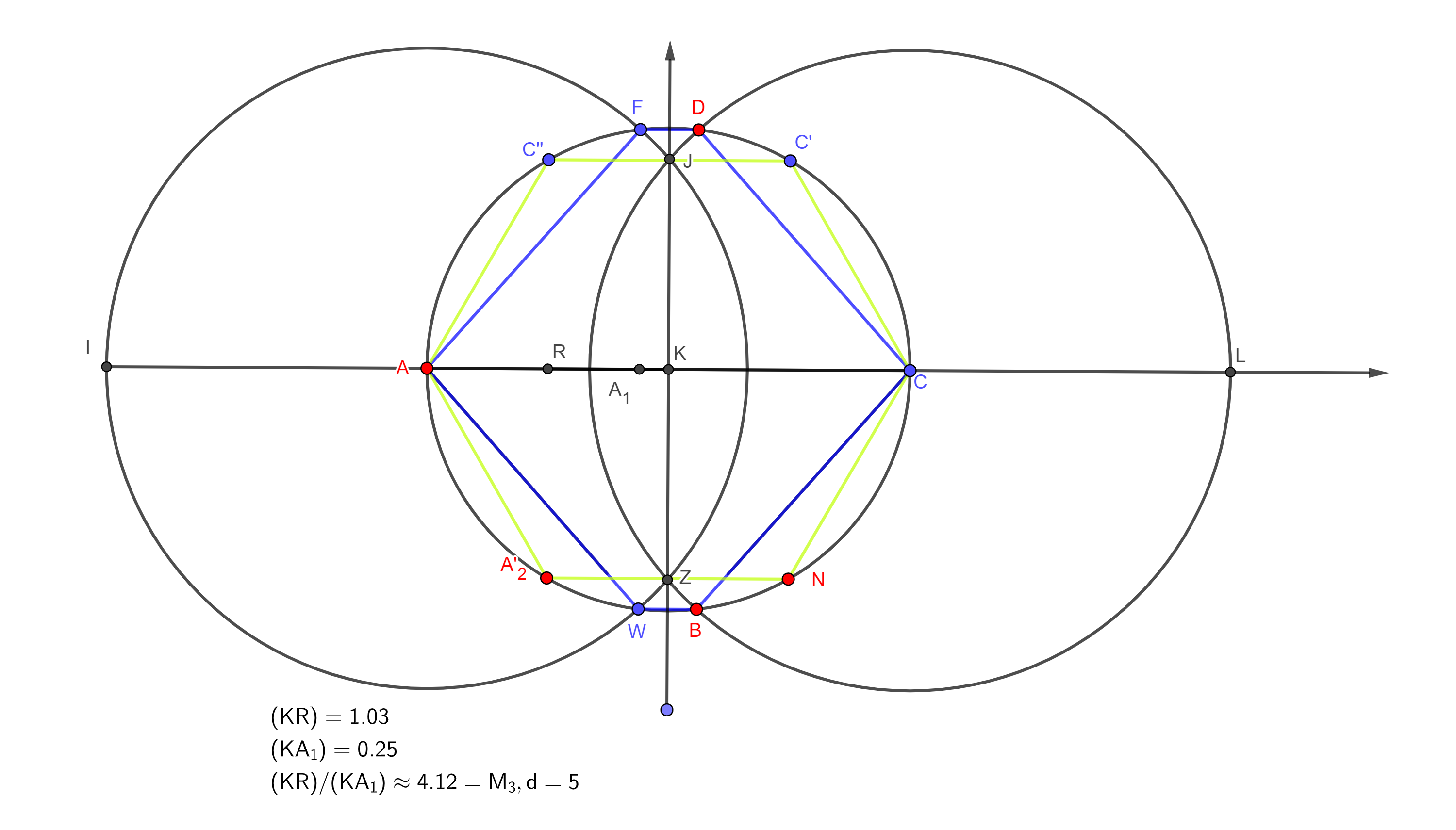}
		\caption{Hexagon conjecture analogue $a*=7\pi/13$}
	\end{figure} 	
		
\begin{figure}
		\centering
		\includegraphics[scale=0.66]{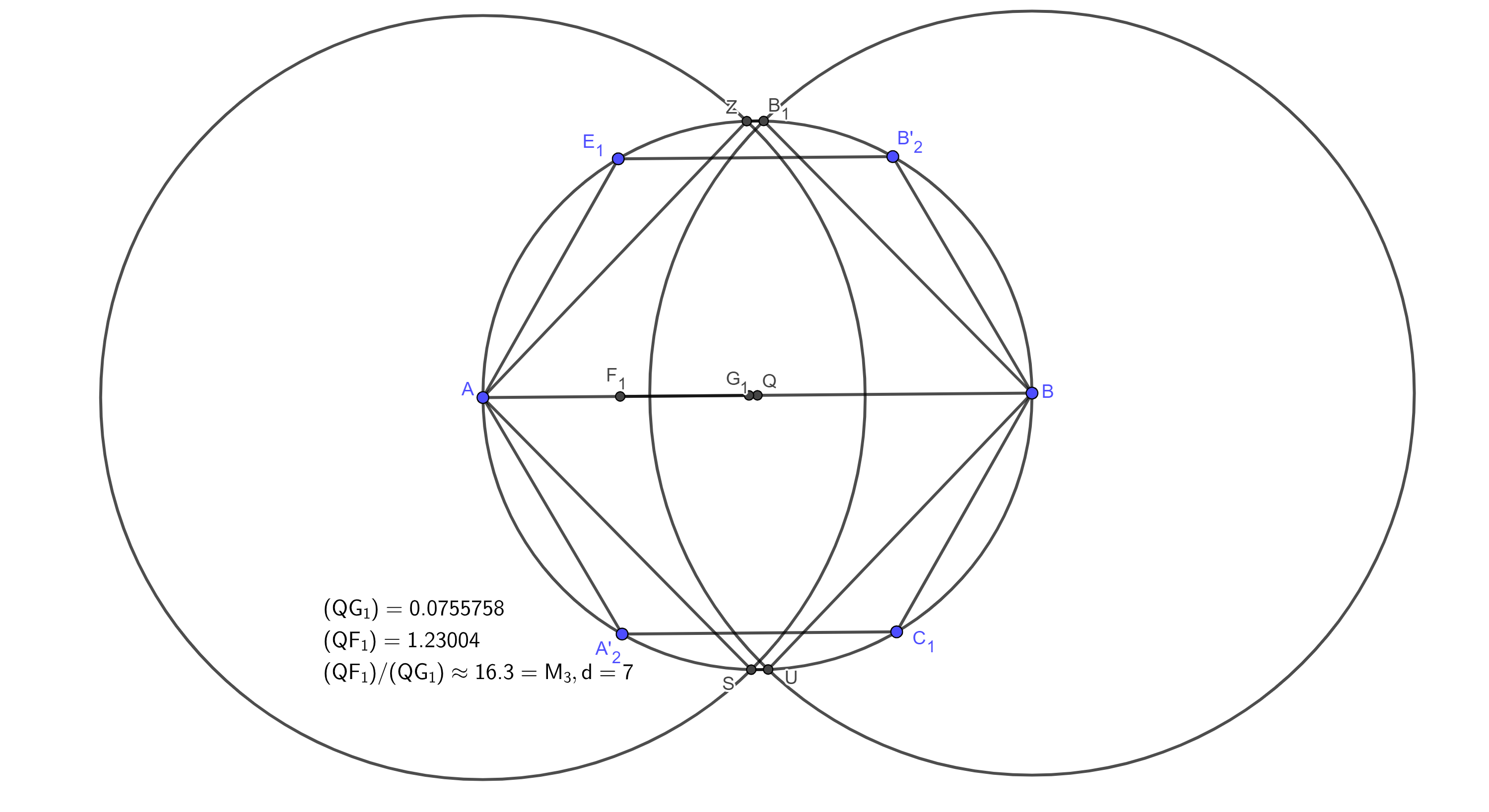}
		\caption{Hexagon conjecture analogue $a*=26\pi/51$}
	\end{figure} 	

\begin{figure}
		\centering
		\includegraphics[scale=0.66]{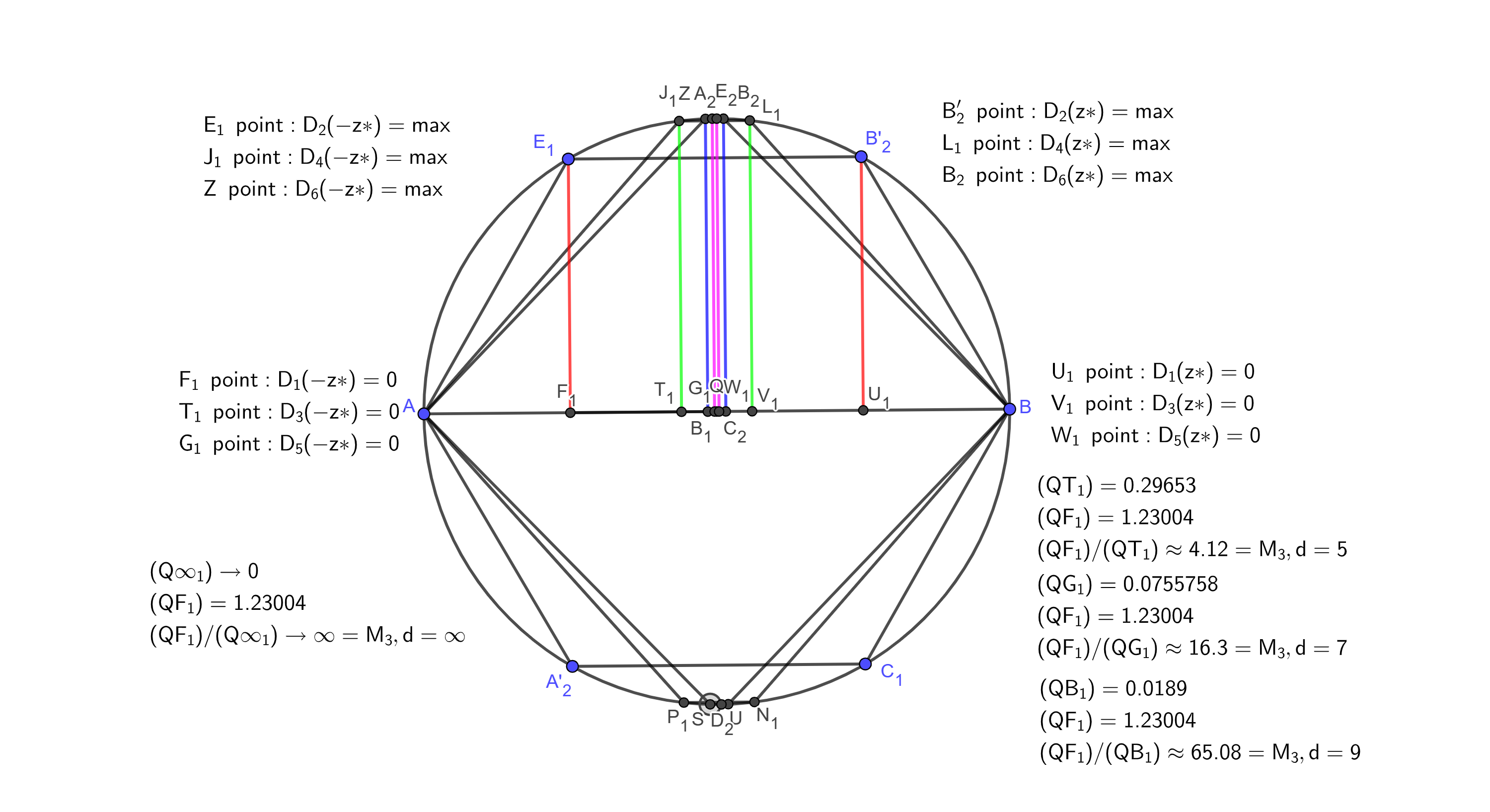}
		\caption{Hexagon conjecture}
	\end{figure}

	\section{Summary and discussion}
	
	The fermion-boson duality at imaginary chemical potential enables me to relate the partition functions of the fermionic $U(N)$ Gross-Neveu model and the bosonic CP$^{N-1}$ model at finite temperature to the identity of complex conjugate Bloch wave functionals that travel inside a periodic environment. These waves are scattered by lattices at angles that are related to characteristic values of the chemical potential and describe the phase transitions of the models but also the transformations of hexagonal and triangular lattices to square lattice. The total  transfer momentum from this scattering coincides with the transfer momentum from a generalised Bragg Law where we replace $sin(\theta)$ with $Cl_m(\theta)$. This approach allowed me to use a kind of lattice regularization of the fermionic model at $d>3$ to see how the lower dimensional theories behave when the highest odd dimension is at its critical point (figure 13).As the dimension increases, the thermal windows within the fermionic and bosonic models are subject to phase transformations, have the boundaries at $\alpha*=\pi/2$ and $\alpha*=3\pi/2$ where now the fermions for $\alpha*=\pi$ and the bosons for $\alpha*=0$ have infinite masses (at strong coupling for the fermions and weak coupling for the bosons). One may find specific values of the mass scales that separate the areas of strong and weak coupling by using the hexagon  (or quadrilateral) conjecture. The projection of the lattice points of the hexagon conjecture onto the real axis create some segments which show as how deep we are at the strong coupling regime (for fermions).It would be interesting to understand these ideas better in the future and to examine if there is a limit that my conjectures coincide to the already existing theories of lattice regularization. 

% Itemized list
%\begin{itemize}
%  \item 
%  \item 
%  \item 
%\end{itemize}

% Numbered list
%\begin{enumerate}
	%\item 
	%\item 
	%\item 
%\end{enumerate}

% Bibliography style - if using a .bib file
	%\bibliographystyle{hindawi_bib_style}
	%\bibliography{<bib file name>} % without .bib extension

\begin{thebibliography}{00}
\bibitem{MAFIADOC}
An earlier, straightly related, version of the manuscript with title "The fermion-boson map for large $d$" was presented in MAFIADOC. This new work extends the ideas of the old paper.
\end{thebibliography}
  
  %Or
  
\section*{Acknowledgements}
I would like to thank A. C. Petkou for his useful comments and help. It would be useful to mention that my earlier publication with the title "The fermion-boson map for large d" which is related to the present work was presented in MAFIADOC.

\appendix

\section{The Bloch-Wigner-Ramakrishnan functions $D_d(z)$}
From the usual analytic continuation of the polylogarithms
\begin{equation}
Li_d(z)=\sum_{n=1}^\infty\frac{z^n}{n^d}\,,\,\,\,z\in{\mathbb C}\setminus [1,\infty)\,,\,\,\,d=1,2,3,..\,.
\end{equation}
one can define the following Bloch-Wigner-Ramakrishnan functions \cite{Zagier1,Zagier2} as
\begin{equation}
\label{Ds}
D_d(z)=\Re\left(i^{d+1}\left[\sum_{k=1}^d\frac{(-\ln|z|)^{d-k}}{(d-k)!}Li_k(z)-\frac{(-\ln|z|)^d}{2d!}\right]\right)
\end{equation}
These are real functions of complex variable, analytic in ${\mathbb C}\setminus \{0,1\}$. The functions $D_1(z)$ and $D_2(z)$ - the latter being the original Bloch-Wigner function - are given by
\begin{equation}
\label{D12}
D_1(z)=\Re[\ln(1-z)]-\frac{1}{2}\ln|z|\,,\,\,\,\,\,\,\,\,
D_2(z)=\Im[Li_2(z)]+\ln|z|{\rm Arg}(1-z)
\end{equation}
In the text we used the following properties of $D_d(z)$'s.
\begin{align}
\label{Dd1}
D_d(1/z)&=(-1)^{d-1}D_d(z)\\
\label{Dd2}
 \frac{\partial}{\partial z}D_d(z)&=\frac{i}{2z}\left(D_{d-1}(z)+\frac{i}{2}\frac{(-i\ln|z|)^{d-1}}{(d-1)!}\frac{1+z}{1-z}\right)
 \end{align}
On the unit circle  we have
\begin{align}
\label{Dodd}
&D_{2n-1}(e^{-i\theta})=(-1)^n\Re[Li_{2n-1}(e^{-i\theta})]=(-1)^n Cl_{2n-1}(\theta)\,,\\
\label{Deven}
&D_{2n}(e^{-i\theta})=(-1)^{n+1}\Im[Li_{2n}(e^{-i\theta})]=(-1)^{n}Cl_{2n}(\theta)\,
\end{align}
for $n=1,2,3,..$. 
The Clausen functions $Cl_m(\theta)$ are defined as
\begin{equation}
\label{Clausen2}
Cl_{2n-1}(\theta)\equiv\sum_{k=1}^{\infty}\frac{\cos k\theta}{k^{2n-1}}\,,\,\,\,Cl_{2n}(\theta)\equiv \sum_{k=1}^{\infty}\frac{\sin k\theta}{k^{2n}}\,,\,\,\,n=1,2,..
\end{equation}
and hence they are respectively even/odd functions of $\theta$. For example
\begin{align}
\label{Clausen1}
&D_1(e^{-i\theta})=Cl_1(-\theta)=-\ln|2\sin(\theta/2)|\,,\,\,\,D_2(e^{-i\theta})=Cl_2(-\theta)=-Cl_2(\theta)\\
&D_{2n-1}(e^{i\theta})=Cl_{2n-1}(-\theta)=Cl_{2n-1}(\theta)\,,\,\,\,D_{2n}(e^{-i\theta})=Cl_{2n}(-\theta)=-Cl_{2n}(\theta)\,.
\end{align}

\end{document}